\documentclass[fleqn,usenatbib]{mnras}

\usepackage{newtxtext,newtxmath}
\usepackage[T1]{fontenc}
\usepackage{ae,aecompl}
\usepackage{graphicx}	
\usepackage{amsmath}	
\usepackage{amssymb}	
\usepackage{scalefnt}
\usepackage{threeparttable, tablefootnote}
\usepackage{wrapfig}	
\usepackage{color, soul}	
\usepackage{verbatim}	
\usepackage{subfigure}
\usepackage{gensymb}
\usepackage{pifont} 
\usepackage{lineno} 

\def\code#1{\texttt{#1}}

\newcommand{\grt}{GM/c^3}

\newcommand{\eg}[1]{(e.g. \citealt{#1})}

\newcommand{\orcid}[1]{\href{#1}{\includegraphics[scale=0.04]{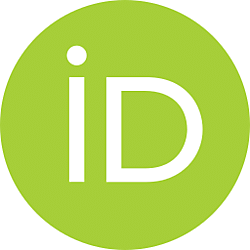}}}

\title[Winds and Feedback from supermassive black holes]{Winds and feedback from supermassive black holes accreting at low rates: Hydrodynamical treatment}

\author[Almeida \& Nemmen]{
Ivan Almeida$^{1}$\orcid{https://orcid.org/0000-0001-6018-2852} \thanks{E-mail: ivan.almeida@usp.br}
and Rodrigo Nemmen$^{1}$\orcid{https://orcid.org/0000-0003-3956-0331} 
\\
$^{1}$Universidade de S\~ao Paulo, Instituto de Astronomia, Geof\'{\i}sica e Ci\^encias Atmosf\'ericas, Departamento de Astronomia,\\ S\~ao Paulo, SP 05508-090, Brazil
}

\date{Accepted 2020 January 03. Received 2019 December 10; in original form 2019 May 29. }

\pubyear{2020}

\begin{document}
\label{firstpage}
\pagerange{\pageref{firstpage}--\pageref{lastpage}}
\maketitle

\begin{abstract}
Outflows produced by a supermassive black hole (SMBH) can have important feedback effects in its host galaxy. An unresolved question is the nature and properties of winds from SMBHs accreting at low rates in low-luminosity active galactic nuclei (LLAGNs). We performed two-dimensional numerical, hydrodynamical simulations of radiatively inefficient accretion flows onto nonspinning black holes. We explored a diversity of initial conditions in terms of rotation curves and viscous shear stress prescriptions, and evolved our models for very long durations of up to $8 \times 10^5 GM/c^3$. Our models resulted in powerful subrelativistic, thermally-driven winds originated from the corona of the accretion flow at distances $10-100 GM/c^2$ from the SMBH. The winds reached velocities of up to $0.01 c$ with kinetic powers corresponding to $0.1-1 \%$ of the rest-mass energy associated with inflowing gas at large distances, in good agreement with models of the ``radio mode'' of AGN feedback. The properties of our simulated outflows are in broad agreement with observations of winds in quiescent galaxies that host LLAGNs, which are capable of heating ambient gas and suppressing star formation.
\end{abstract}


\section{Introduction}\label{sec:introduction}

When matter falls into a black hole (BH) it forms a disk-like structure due to the barrier posed by angular momentum conservation--an accretion flow. Magnetic stresses in the ionized plasma introduce friction which allows the gas to flow in toward the BH \citep{Balbus2003}. At the same time, these stresses convert some of the gravitational potential energy of the accretion flow into heat and can release a substantial fraction of its rest mass energy, providing the primary power source behind active galactic nuclei (AGNs), black hole binaries and gamma-ray bursts \citep{Meier2012}.

The dynamics of the resulting accretion flow depends critically on whether the viscously generated thermal energy is radiated away \citep{Abramowicz2013}. This is parameterized in terms of the radiative efficiency $\epsilon = L/\dot{M} c^2$ where $L$ is the luminosity produced by the accretion flow and $\dot{M}$ is the mass accretion rate onto the BH. In this paper, we are particularly interested in the regime of BHs accreting at low $\dot{M}$. At rates $\dot{M} \lesssim 0.01 \dot{M}_{\rm Edd}$ ($\dot{M}_{\rm Edd}$ is the Eddington accretion rate), the gas cannot radiate its thermal energy away and becomes extremely hot ($T \sim 10^{12}$ K), geometrically thick ($H \sim R$, $H$ is the vertical disk thickness) and optically thin, giving rise to a radiatively inefficient accretion flow (RIAF) with $\epsilon \ll 1$ \citep{Yuan2014}. The sheer majority of SMBHs in the local universe--inactive galaxies and low-luminosity AGNs (LLAGNs)--are fed at low, sub-Eddington rates and hence in the RIAF mode, with the nearest example being Sagittarius A* (Sgr A*), the $4 \times 10^6 \, M_\odot$ BH at the center of Our Galaxy \citep{Narayan1995nat,Yuan2003}. 

The presence of a SMBH accreting in the RIAF mode can have important feedback effects in its host galaxy. In the centers of many galaxy clusters the ``radio mode'' of feedback has been observed in the form of powerful radio jets heating the cluster atmospheres and offsetting cooling \eg{McNamara2012}; these clusters usually host a SMBH accreting at low $\dot{M}$ \eg{Birzan2004, Nemmen2015}. 
There is also evidence for feedback operating in individual galaxies in the form of centrally driven winds from SMBHs in LLAGNs lacking obvious extended radio jets, dubbed ``red geysers''; these winds carry out enough mechanical energy to heat ambient, cooler gas and thereby suppress star formation \citep{Cheung2016,Roy2018}. 
In fact, it has been proposed that outflows from SMBHs accreting at low rates may be responsible for quenching star formation \citep{Croton2006,Bower2006,Bower2017} and therefore explain the increase in the number of quiescent galaxies--the vast majority of galaxies which have little or no ongoing star formation--over the past ten billion years \citep{Bell2004, Bundy2006, Faber2007, Ilbert2010}. 
Moving on closer to home, a major surprise from the \textit{Fermi} Large Area Telescope was the detection of the \textit{Fermi} bubbles above and below the direction of the galactic center \citep{Su2010,Ackermann2014a}. 
One possibility is that the SMBH at the center of the Milky Way may once have had a stronger activity at its nucleus like that of a brighter AGN, producing powerful outflows within the past few million years \citep{Guo2012, Mou2014}. 
It is clear that properly modeling RIAFs and their outflows is relevant for the full understanding of AGN feedback.

There is a considerable body of work on the theory of RIAFs. Here, we briefly summarize the progress focusing on numerical simulations of wind launching from RIAFs. The early work focused on deriving analytical one-dimensional solutions to the RIAF structure \citep{Narayan1994, Narayan1995b}; they suggested that the positivity of the Bernoulli parameter in the solutions implies that the gas is weakly bound to the BH --$Be$ is defined as

\begin{equation}
Be = \frac{v^2}{2} + \gamma \frac{P}{\rho} + \psi.
\label{Be-eq}
\end{equation}
where $v$, $P$, $\rho$ and $\psi$ are, respectively, velocity, pressure, density and gravitational potential. Therefore, RIAFs would be quite likely to produce outflows.  \cite{Blandford1999, Begelman2012} took the argument to the extreme, suggesting the RIAFs are always accompanied by vigorous outflows and proposed the ansatz that  the inflow rate follows $\dot{M}(r) \propto r^s$--i.e. a reduction in the inflow rate due to mass-loss in winds. \cite{Abramowicz2000} argued that the Bernoulli parameter is irrelevant to judge whether outflows are produced by the system but pointed out that RIAFs may have--but do not need to have--winds.

While analytical one-dimensional models are very useful, some aspects of accretion physics such as the formation of outflows and their nonlinear dynamics are beyond the scope of such models. Numerical simulations are needed to properly model these systems. The first global simulations of RIAFs were purely hydrodynamic and Newtonian \citep{Stone1999, Igumenshchev1999, Igumenshchev2000}. They found that the accretion flows are convective and observed strong bipolar outflows. \cite{Proga2003a} used a pseudo-Newtonian potential and ignored viscosity; Proga \& Begelman found no outflows in their work. More recently, \cite{Yuan2012, Yuan2012b, Bu2016a} performed hydrodynamic simulations of RIAFs with an increased dynamical range encompassing from near the Bondi radius down to the BH . They found fairly strong outflows and an apparent support to the $\dot{M}(r)$ ansatz of \cite{Blandford1999}. 
\cite{Li2013, Bu2018, Bu2018b} included a cooling term in the energy equation and found strong, thermally-driven winds. 

The next step of numerical work consisted of advancing beyond hydrodynamic models and adding magnetic fields in order to explore the magnetorotational turbulence and the effect of different initial configurations of magnetic fields on the disk and wind evolution. \cite{Machida2000, Machida2001} performed global magnetohydrodynamic (MHD) simulations of RIAFs and found the development of temporary outflows. Similarly, \cite{Igumenshchev2003} found an initial transient bipolar outflow, however in the latter work the transient is followed by a steady state weak thermal wind. \cite{Stone2001, Hawley2002, Proga2003, Bu2016} observed strong outflows at all radii beyond the innermost stable circular orbit in their MHD models. The MHD simulations of \cite{Pen2003, Pang2011} showed no sign of outflows. 

\cite{DeVilliers2003} inaugurated the era of global, general relativistic MHD (GRMHD) simulations of RIAFs. \cite{DeVilliers2003, DeVilliers2005} observed two types of outflows in their models: relativistic, Poynting-flux dominated jets along the poles of the BH and a coronal matter-dominated wind that did not have enough energy to escape to infinity and hence was bound to the BH (cf. also \citealt{McKinney2004, Hawley2006}). \cite{Sasha2012a, McKinney2012} performed GRMHD simulations of larger tori with an emphasis on understanding the dynamics of jets. They found relatively strong, magnetized winds with a power depending on the BH spin and carrying as much as $\approx 10\%$ of the rest-mass energy associated with accreted matter to infinity, similarly to \cite{Sadowski2013,Sadowski2016}. The simulations of \cite{Moscibrodzka2013, Moscibrodzka2014, Moscibrodzka2016a} also find magnetized coronal winds, though they do not quantify the energy carried by such outflows. Puzzlingly, \cite{Narayan2012} found little evidence for winds in their GRMHD models with large tori, long durations and  different magnetic topologies. Narayan et al. pointed out that the limited convergence of their models prevents them from drawing more robust conclusions on the amount of mass-loss in winds from RIAFs. Interestingly enough, \cite{Yuan2015} reanalyzed the simulation data of \cite{Narayan2012} using Lagrangian particles and found winds that carry $\sim 1\%$ of the rest-mass energy associated with accreted matter to infinity. 

From the literature review presented above, it is clear that the issue of wind-launching from RIAFs is not settled. Some of the unresolved questions are: do the winds produced by underfed SMBHs provide significant feedback inside the host galaxy? In other words, do they carry enough energy and momentum to be able to heat up gas, shut down star formation and therefore impact the evolution of galaxies? What are the energy, momentum and mass outflow rates from such systems?  These are the main broad questions that this paper will address. 

This work employs numerical simulations for studying the global, multidimensional physics of hot accretion flows. More specifically, here we perform global two-dimensional hydrodynamical simulations of RIAFs around non-spinning BHs, with the goal of investigating in a self-consistent way the winds produced by accreting SMBHs such as those that inhabit the centers of nearby galaxies, and the possible feedback effects in their environment. 
Since we wanted to keep the simulation conditions as general as possible, we considered only a Schwarzschild BH and did not assume initial conditions with particular magnetic topologies (such as e.g. \citealt{Narayan2012}), keeping the simulation purely hydrodynamic. Because the BHs in our models are not spinning, we will not have energy extraction from Kerr spacetime and hence no Blandford-Znajek driven polar jets \citep{Blandford1977}. This is by design, since we know that jets occur in only $\approx 10\%$ of AGNs \citep{Kellermann1989}--therefore they cannot account for AGN feedback in the vast majority of quiescent galaxies--and they are also collimated and therefore may not interact efficiently with the interstellar medium.

Technically, the novelty of this work compared to many previous numerical simulations of hot accretion flows in the literature is the following:
(i) some of our models are the longest running simulations of RIAFs so far produced, with durations of up to $8 \times 10^5 \grt$; 
(ii) our models have a large dynamical range, with the initial outer edge of the torus extending to $500 R_S$;
(iii) we explored a prescription for viscous stress tensor based on GRMHD simulations \citep{Penna2013b};
(iv) in some of our models, we adopted the equilibrium torus solution of \cite{Penna2013}, which corresponds to a more physical initial condition than earlier torus solutions;
(v) finally, we used a Lagrangian tracer particles to improve the estimates of quantities associated with the outflows. 

The structure of this paper is as follows. In section \ref{sec:methods} we outline the computational methods used to solve the fluid conservation equations, initial conditions, parameter space and techniques used in the analysis. In section \ref{sec:results} we describe the results: the temporal evolution of the flow, amount of energy, momentum and mass outflow rates, geometry, collimation and launching radii of winds, and the density profile of the accretion flow. In section \ref{sec:disc} we contextualize our results, comparing our simulated accretion flows and outflows with  observations of LLAGNs and AGN feedback, and also with previous numerical models. Finally, we conclude with a summary and perspectives in section \ref{sec:end}. 
Readers interested in the density profiles of the hot accretion flow can skip to section \ref{subsec:accretion-flow}. Those interested in the outflow properties and feedback efficiency should go to sections \ref{subsec:eff}, \ref{subsec:lag-part}. The comparisons with observations of LLAGNs, Sgr A* and AGN feedback can be found in section \ref{subsec:obs}.

\section{Methods}\label{sec:methods}

\subsection{Computational method} \label{subsec:comp-method}

In this work, we aim at simulating the evolution of thick accretion flows around black holes. We are particularly interested in understanding the origin and development of subrelativistic winds from black holes, for which the extraction of spin energy from the black hole is thought to be not so important--as opposed to relativistic jets. For this reason, we considered only a Schwarzschild black hole and adopted a Newtonian hydrodynamical (HD) treatment, describing the black hole gravity in terms of the pseudo Newtonian potential (\citealt{Paczynsky1980}; cf. section \ref{subsec:equations}). 

We performed our numerical simulations with the Eulerian \code{PLUTO} code\footnote{We used version 4.2 of the \code{PLUTO} code, commit \code{8ffd30330ecf91f08ca6cda5f9e61492cae55e3e} available at \url{https://github.com/black-hole-group/pluto}.} which solves the hyperbolic system of conservation equations of Newtonian fluid hydrodynamics  using the finite volume approach based on a Godunov-type scheme \citep{Mignone2007}. We did not take into account electromagnetic fields explicitly; instead, we incorporated the energy and angular momentum dissipation expected due to the magnetorotational instability (MRI; \citealt{Balbus2003}) by means of an appropriate viscous stress tensor (cf. section \ref{subsec:equations}). 

We adopted units such that $GM = 1$ and the Schwarzschild radius is unitary, $R_S \equiv 2GM/c^2 = 1$ (i.e. $c= \sqrt{2}$). Length and time in this paper are given in units of $R_S$ and $GM/c^3$, respectively. $R$ corresponds to the radius in cylindrical coordinates, $r$ in spherical ones.

Our simulations run for a very long time, since we are interested in the \textit{global} dynamics of the accretion flow and winds. We can make a rough estimate of the simulation duration necessary for the flow state to converge. The basic idea is that we expect the flow to reach a steady state equilibrium on a timescale comparable to the viscous time $t_{\rm visc}$. The simple self-similar ADAF model \citep{Narayan1994} gives us useful scalings, according to which the viscous time at a radius $R$ is given by 
\begin{equation}
t_{\rm visc} = \frac{r}{v_r} \sim \frac{t_{\rm ff}}{0.5 \alpha}
\end{equation}
where $\alpha$ is the Shakura-Sunyaev viscosity parameter and $t_{\rm ff}$ is the free-fall timescale. This simple model indicates that in order for a parcel of gas located at $r=500 R_S$ in the disk to achieve inflow equilibrium, it would take an amount of time $t \sim t_{\rm visc} = 2 \times 10^5 \grt$ for $\alpha=0.3$. Therefore, our simulations need to have a comparable duration in order to ensure that the flow achieves inflow equilibrium (i.e. convergence) in at least part of the domain, thus justifying the long running times. The running time of the simulations varied between $4 \times 10^4$ to $8 \times 10^5 \grt$, depending on whether we found a specific simulation to be more promising in terms of its potential for wind launching potential. An analysis of the radial extension within which the models attained inflow equilibrium is presented in appendix \ref{app:convergence}. 

Our black hole accretion flow simulations have the longest duration to date, to our knowledge. The long durations of our models imply that they are usually quite computationally expensive. For this reason, we have chosen to restrict the dimensionality of our models to only two dimensions.

\subsection{Equations set}\label{subsec:equations}

The set of equations describing hydrodynamic accretion flows were presented in \cite{Stone1999}; these equations are reproduced below: 
\begin{align}
& \frac{d\rho}{dt} + \rho {\nabla} \cdot \mathbf{v} = 0\text{,} \label{mass-conservation} \\
& \rho\frac{d\mathbf{v}}{dt} = {\nabla}P - \rho {\nabla} \psi + {\nabla} \cdot \mathbf{T}\text{,} \label{momentum-conservation} \\
& \rho\frac{d\mathbf{e/\rho}}{dt} = -P{\nabla} \cdot \mathbf{v} + \frac{\mathbf{T}^2}{\mu}\text{.}
\label{energy-conservation}
\end{align}
In equations \eqref{mass-conservation} - \eqref{energy-conservation}, $\rho$ is the density, $\mathbf{v}$ is the velocity, $P$ is the pressure, $e$ is the internal energy density, $\psi$ is the gravitational potential and $\mathbf{T}$ is the anomalous stress tensor. We adopted the pseudo Newtonian potential $\psi = GM/(r-R_S)$ which incorporates the basic features of the Schwarzschild geometry \citep{Paczynsky1980}.

In order to incorporate angular momentum transport that mimics MRI, we followed \cite{Stone1999} and assumed that the non-azimuthal components of $\mathbf{T}$ are zero; the non-zero terms of $\mathbf{T}$ are, in spherical-polar coordinates ($r, \theta, \phi$):
\begin{align}
& T_{r\phi} = \mu r \frac{\partial}{\partial r}\left( \frac{v_{\phi}}{r} \right)\text{,} \label{T-radial} \\
& T_{\theta\phi} = \frac{\mu \sin \theta}{r} \frac{\partial}{\partial \theta}\left( \frac{v_{\phi}}{\sin \theta} \right)\text{,} \label{T-phi}
\end{align}
where $\mu = \nu \rho$ is the viscosity coefficient and $\nu$ is the kinematic viscosity \citep{Landau1959}. For astrophysical systems, the plasma microphysics details are still an open question. The wind production is heavily dependent on how the angular momentum is transferred across the accretion disc. The effective viscosity generated by MRI is not well constrained, depending on the initial magnetic field topology. In this work we explored two different prescriptions for the viscous stress by adopting different parameterizations for $\nu$: 
\begin{enumerate}
\item $\nu = \alpha r^{1/2}$ which corresponds to the ``K-model'' in \cite{Stone1999}. We will refer to this $\nu$-scaling as ST.
\item $\nu = \alpha c_s^2/ \Omega_K$ following \cite{Shakura1973}. We will refer to this parameterization as SS.
\end{enumerate}
In the above expressions, $\Omega_K$ is the Keplerian angular velocity and $c_s$ is sound speed. The $\alpha$ parameter is the usual Shakura-Sunyaev $\alpha$-parameter for accretion discs \citep{Shakura1973} which we allow to vary in the range $0.01 \leq \alpha \leq 0.3$. Note that, strictly speaking, the correspondence between the $\alpha$ here and the ``Shakura-Sunyaev $\alpha$'' is exact only in the SS model. $\alpha$ is related to turbulence scales in the system, by definition in \cite{Shakura1973}. Since $\alpha$ is a multiplicative factor in $\nu$ expression, it is connected to the efficiency of the angular momentum energy transport. This is a traditional parameter explored in the literature of accretion flow simulations \citep{Stone1999, Yuan2012b}.

We also explored a model in which $\alpha$ varies with radius, i.e. $\alpha=\alpha(r)$, inspired on \cite{Penna2013b}. Penna et al. obtained an analytical approximation to $\alpha(r)$ that reproduces well the numerical GRMHD simulations of RIAFs, which we reproduce here: 
\begin{equation}
\begin{aligned}
\alpha(r) = 
\begin{cases}
     \frac{1}{40} \left( \frac{1-\frac{1}{r}}{1 - \frac{1.5}{r}}\right)^6,            & r > 3R_S \\
     0.140466,  & r < 3R_S
\end{cases}
\end{aligned}
.
\label{alpha-penna}
\end{equation}

\subsection{Initial conditions and grid}\label{subsec:init-cond}

Our initial condition consists of a rotating HD torus in dynamical equilibrium with a specific angular momentum profile $l(R)$. The torus' inner edge is located at $R_{\rm in} = 5-20R_S$--this range is due to numerical reasons and the specific choice of $R_{\rm in}$ depends on $l(R)$--and outer edge $R_{\rm out} \approx 500 R_S$. The radius of maximum density $R_0$ was varied in our models in the range $R_0 \approx 10-25 R_S$ depending on the $l(R)$ model adopted, bound by the values of $R_{\rm in}$ and $R_{\rm out}$. Our torus is pretty large--larger than most simulations which usually begin with a torus ending at $\approx 40 R_S$ (e.g., \citealt{Moscibrodzka2013, Porth2017})--since we are interested in both the density profile up to larger scales and whether winds are launched at larger radii from the disk. In this work we defined the total torus mass $M_0$ as $M_0 = \int \rho(\mathbf{r}, t=0)dV$, with the following normalization: $\max (\rho) = 1$. For our system $M_{\rm BH} \gg M_0$, so we neglected any effects from torus self-gravity.

The $l(R)$ is responsible to describe the rotation of the initial torus. In this sense, this parameter describes how the kinetic energy was stored inside our initial system -- since we had the stationary initial condition imposed by $v_r(t=0) = 0$ and $v_z(t=0)-0$. We explored two $l(R)$-profiles in our simulations, both depending only on the cylindrical radius $R$: 
\begin{enumerate}
\item Power-law scaling $l(R) \propto R^a$, where $0 \leq a < 0.5$. \cite{Papaloizou1984} reported a full analysis of the $a=0$ case. Here, we considered three different values of $a$: 0.0, 0.2 and 0.4.
\item $l(R)$ piecewise scaling proposed by \cite{Penna2013}, adapted to a non-relativistic torus:  $l={\rm constant}$ for $R<21R_S$, $l(R)=0.71 l_K$ elsewhere where $l_K$ is the Keplerian specific angular momentum.
\end{enumerate}

For the Power-law profile, the value of $a$ is limited to $a<0.5$. For $a=0.5$ one would have an infinitely thin disc. Increasing the value of $a$ leads to thinner initial discs. Previous works \citep{Stone1999, Yuan2012} used discs with $a=0$ to initialize the RIAF. However, the initial state of SMBH accretion is unknown such as the velocity field in the innermost regions. For this reason, we explored different $l(R)$ profiles in our simulation sample. Intermediary values of $a$ were allowed in order to understand the influence of rotation in RIAFs (all our profiles presented sub-keplerian motion). In addition to the power-laws, we studied the profile proposed by \cite{Penna2013} which is a initial condition used in the RIAF simulations literature \citep{Narayan2012}. The choice of $l(R)$ is important to the energy balance of the system and the total energy initially available. Higher values of angular momentum lead to higher values of the Bernoulli parameter. High rotation is naturally a more fertile terrain for wind production, since it is easier to produce $Be > 0$. We are interested in any feature that can modify the outflows dynamic as iti is known that different initial setups for rotation can lead to different results for accretion flow simulations \citep{Bu2013}.

\begin{figure*}
\vspace{\fill}
\noindent
\makebox[\textwidth]{
\subfigure[][$l(R) = $ constant]{\includegraphics[width=\linewidth/4]{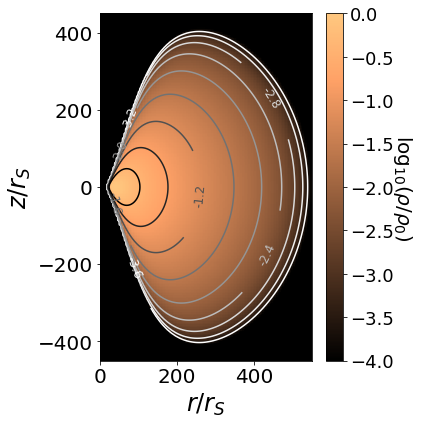}}
\subfigure[][$l(R) \propto R^{0.2}$]{\includegraphics[width=\linewidth/4]{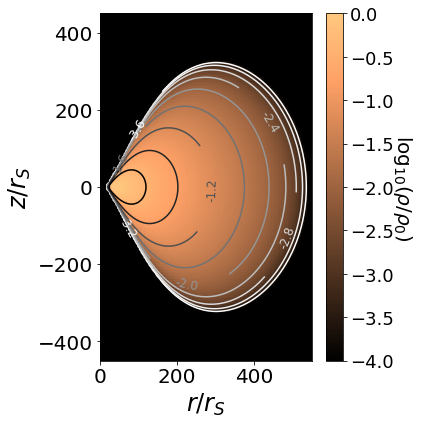}}
\subfigure[][$l(R) \propto R^{0.4}$]{\includegraphics[width=\linewidth/4]{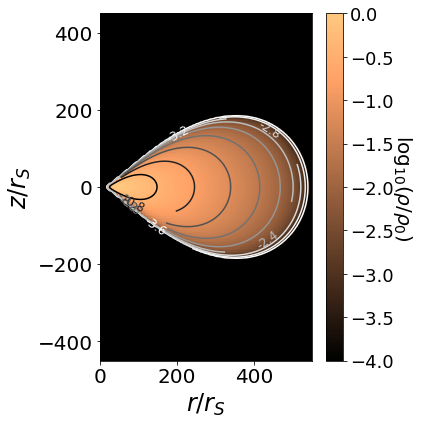}}
\subfigure[][$l(R)$ inspired on \cite{Penna2013}]{\includegraphics[width=\linewidth/4]{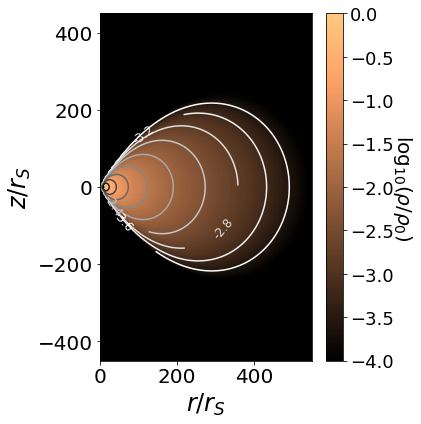}}
}
\caption{Torus density distribution for the four specific angular momentum profiles considered.}
\label{fig:initial-conditions}
\vspace{1cm}
\end{figure*}

The four torus described above are shown in Figure \ref{fig:initial-conditions}. As can be seen in this figure, $a$ has the effect of changing the torus thickness. The reason why we explored models with $a>0$ is because we wanted to initialize models with a torus thickness $H \sim R$ as expected for RIAF models \citep{Yuan2014}, where $H$ is the scale height.

Regarding the computational domain, we used a fixed mesh and our grid extends to a large radius, $10^4 R_S$--which is one order of magnitude larger than the outer radius of the disc size--in order to avoid undesirable boundary effects. Our grid is uniformly distributed in $\log_{10}$(radius) with 400 cells; as such, the inner regions have a higher resolution. The radius of the computational domain begins at $1.25 R_S$. We adopt the outflow boundary condition at the inner and outer radii.

\begin{figure}
\center
\includegraphics[width=\linewidth]{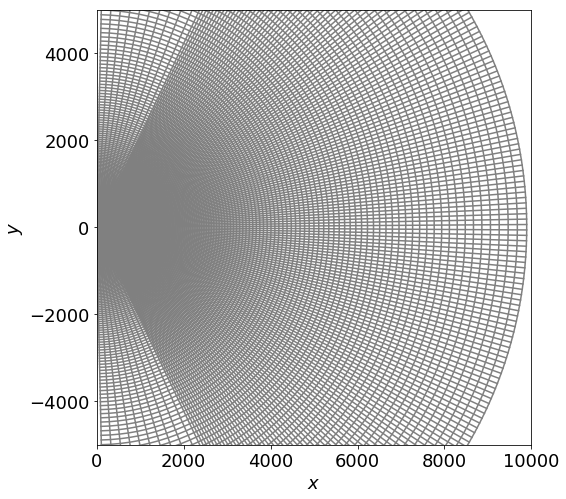}
\caption{Grid used in the simulations.}
\label{fig:grid}
\end{figure}

To avoid numerical errors, the grid is restricted to $2 \degree \leq \theta \leq 178 \degree$. In the $\theta$-direction, we defined two regions with a different number of cells in each, such that we have less cells near the grid poles (Figure \ref{fig:grid}). The regions are separated according to the values of $\theta$:
\begin{equation}
N_{\rm cells}\ {\rm in}\ \theta{\rm -direction} = 
\begin{cases}
10, & \text{if }  \theta < 15^{\circ}  \ \text{or} \ \theta > 165^{\circ} \\
180,  & \text{if} \ 15^{\circ} \leq \theta \leq 165^{\circ}
\end{cases}
.
\end{equation}
The reason why we decreased the spatial resolution near the poles is because we do not expect any significant action to occur in this region. Therefore we have chosen to concentrate the resolution in the polar regions where we expect the development of the accretion flow and wind. If we were simulating the flow around a Kerr BH, then we would expect to have a Poynting flux-dominated jet which would fill the polar regions. However, since we are dealing with a Schwarzschild black hole, our grid choice is appropriate.  
 
\subsection{Lagrangian particle tracking}\label{subsec:traj-app}

One technique that we used to identify and characterize outflows--in addition to analyzing the evolution of the mass and energy fluxes across our mesh-based simulations--was to introduce ``tracer'' particles which are be passively advected with the fluid flow, and thereby track its Lagrangian evolution, allowing the thermodynamical history of individual fluid elements to be recorded. This technique is called Lagrangian particle tracking and has been used to make sense of several astrophysical simulations (e.g. \citep{Enslin2002, Dubois2012, Genel2013, Yuan2015}). It is particularly useful in our simulations, since it does not rely on using the Bernoulli parameter which is an indirect way of assessing whether outflows were produced, therefore being a more appropriate outflow measure. 

We implemented the traditional scheme in which the tracer particles are massless particles advected in space using the local velocity field \citep{Harlow1965}. To obtain the trajectories of the particles, we solve the differential equation
\begin{equation}
\frac{d \mathbf{x_p}}{dt} = \mathbf{v_f}(\mathbf{x_p}, t)
\end{equation}
where $\mathbf{x_p}(t)$ is the particle position and $\mathbf{v}$ is the fluid velocity at the position $\mathbf{x_p}$. With the velocities from simulation data at a particular time $t$, we can advance the position of the tracer particle to $t + \Delta t$ which is accurate to first-order, limited by the time-resolution of the simulation. 

The simulations' time step $\Delta t$ were chosen to be sufficiently short--approximately the orbital Keplerian period $t_K$ at $R \approx 8R_S$--such that the distance a fluid element is able to cover over a timescale $t_K$ is much smaller than the size of the disc, $v \Delta t \ll R_{\rm out}$ where in this context $v$ is a typical fluid velocity. 

In order to assess whether outflows are produced from a given simulation and--in case there is an outflow--to quantify its properties, we used a set of 1250 tracer particles. We started the particle tracking at the moment when the fluid has reached a stationary net mass accretion rate, i.e. when the value of $\dot{M}_{\rm acc}(R_{\rm in}, t)$ (cf. equation \ref{mdot-in}, Figure \ref{fig:acc-time}) becomes roughly constant; we defined this moment as $t_0$. The particles were initially uniformly distributed in the $r-\theta$ space. The particles are separated in $r$ by steps of $4.3R_S$ and in $\theta$ by steps of $6.25\degree$. The region of the particles were delimited by the ranges: $R=40R_S-250R_S$ and $\theta=15\degree - 165\degree$ -- the distribution was with 50 particles over $r$-coordinate and 25 particles over $\theta$-coordinate. For $t>t_0$, we let the particles to be advected by the flow and monitored their positions with time.

In this work we adopted two criteria for identifying whether a tracer particle is part of an outflow. Firstly, since we are only interested in the properties of winds, we reject particles which are located near the poles--the domain of the relativistic jet if we had a Kerr BH--or in the accretion disc. In order to perform this rejection, one straightforward approach is to consider only particles within a limit range of polar angles. Here, we consider as outflowing particles only those which have reached $15\degree \leq \theta \leq 45\degree$ or $135\degree \leq \theta \leq 165\degree$ at the end of the simulation, following the results of \cite{Sadowski2013} who find that subrelativistic winds are limited to a similar range (cf. also \citealt{Moscibrodzka2014}).

Secondly, based on the final radius $r_{\rm final}$ of the particle we have defined two types of outflow:
\begin{enumerate}
\item If $R(t=t_0) < r_{\rm final} < 500R_S$, we call this ``simple outflow'', i.e. the particle was not accreted but also did not reach very far away.
\item If $r_{\rm final} > 500R_S$ we call ``real outflow'', i.e. the particle reaches a distance larger than the maximum radius of the original torus ($r_{\rm final} > R_{\rm out}$).
\end{enumerate}
We adopted this criterion because it is a simple estimator of whether the particle was able to reach a sufficiently large distance from the black hole. In our tests, particles that reach distances of $\sim 500R_S$ usually have $Be>0$---i.e. the particle is likely unbound---since the gravitational binding energy decreases with distance and the kinetic energy and enthalpy dominate over gravitational energy.

Based on the final velocity of the outflowing particles we defined: 
\begin{enumerate}
\item If $v_r > 0$ we considered the particle as a true outflowing particle.
\item If $v_r < 0$, we call these particles as ``fallback particles'', since they were ejected once but eventually returned towards the BH.
\end{enumerate}

\begin{figure}
\noindent
\includegraphics[width=\linewidth]{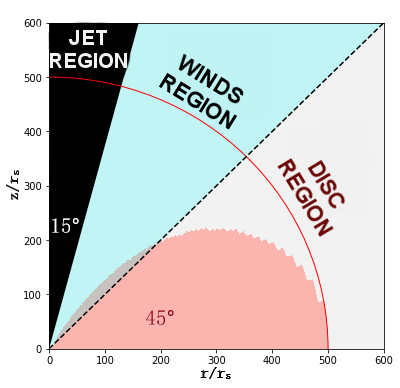}
\caption{Schematic drawing of the different regions of the flow. Jet region are defined as a region near the pole with 15\degree opening, and the disc region is a region near the equator with 45\degree opening; all material ejected in these two regions are excluded in our analysis of outflows because this regions are believed to be dominated by jet and accretion disc, respectively, in nature. We considered only the region between both, that we called wind region and is represented in blue. The red solid line is the outflow limit that we have defined, every material that it is in the wind region and beyond the red line was classified as real outflow. The pink solid region is a representation of our initial torus.}
\label{fig:zones}
\end{figure}

Following these criteria, the ``wind region'' is illustrated in Figure \ref{fig:zones}; particles that get outside the red circle are presumably part of a wind launched by the black hole. Results from GRMHD simulations support the basic aspects of this picture \citep{Sadowski2013, Moscibrodzka2014}

\subsection{Simulation setup}\label{subsec:sim-setup}

We performed a total of 11 simulations exploring the variation of three main properties of the flow: the specific angular moment profile $l(R)$, the viscosity prescription $\nu$ and the value of $\alpha$; the parameter space of simulations is summarized in Table \ref{tab:simulations}. It is important to investigate different $l(R)$-profiles since the actual rotation curve of RIAFs in nature is not known. In particular, we do not know the initial conditions of SMBH accretion in low-luminosity AGNs, and the long-term evolution of the accretion flow and possible winds could be dependent on these initial conditions, which is an incentive to not be too conservative in choosing the parameters of our numerical experiments. 

\begin{table}
\centering
\begin{tabular}{llcccc}
 \textbf{\#ID} &\textbf{Name} & \textbf{$l(R)$} & \textbf{$\nu$} & \textbf{$\alpha$} & \textbf{Duration  $10^5 \left[ \frac{GM}{c^3} \right]$} \\ \hline
00 & PNST.01 & Penna2013        & ST     & 0.01  & 8.0  \\ 
01 & PNST.1 & Penna2013        & ST     & 0.1   &  0.9 \\  
02 & PNSS.1 & Penna2013       & SS     & 0.1   & 4.5 \\ 
03 & PNSS.3 & Penna2013        & SS     & 0.3   & 3.3 \\ 
04 & PNSSV & Penna2013        & SS     & $\alpha (r)$   & 3.8 \\ 
05 & PL0ST.1 & $a=0.0$         & ST     & 0.1   & 0.8  \\ 
06 & PL0SS.3 & $a=0.0$         & SS     & 0.3   & 2.1 \\ 
07 & PL2SS.1 & $a=0.2$         & SS     & 0.1   & 1.4 \\ 
08 & PL2SS.3 & $a=0.2$         & SS     & 0.3   & 2.1 \\ 
09 & PL2ST.1 & $a=0.2$         & ST     & 0.1   & 0.4  \\ 
10 & PL4ST.01 & $a=0.4$         & ST     & 0.01  & 1.7  \\ \hline 
\end{tabular}
\label{tab:simulations}
\caption{List of the numerical simulations performed in this work. The second column refers to the specific angular momentum. ``Penna2013'' refers to the torus described in \citealt{Penna2013} and the others are related to a power-law form $l(R) \propto R^a$ (see section \ref{subsec:init-cond});
$\nu$ and $\alpha$ columns refer to the adopted viscosity profile and the dimensionless coefficient (see \ref{subsec:equations} ).
}
\end{table}

The other two parameters--$\nu$ and $\alpha$--are responsible for the angular momentum transport that allows accretion to proceed. We described the two parameterizations of $\nu$ that we adopted in section\ref{subsec:init-cond}. We expect the long-term behavior of the flow to strongly depend on the functional form of $\nu$. Moreover, $\alpha$ regulates the strength of the angular momentum removal as in the classical Shakura-Sunyaev solution. We chose values of $\alpha$ consistent with estimates from global and shearbox simulations of the MRI process in BH accretion flows (cf. \citealt{Penna2013b} for a review).

As argued in section \ref{subsec:comp-method}, we ran the simulations for a long time--comparable to the viscous time at large radii in the disc--in the hopes that a considerable part of the accretion flow converges. The individual duration of each model was different based on whether we found each interesting in terms of wind production. Models that did not show clear signs of winds were not allowed to develop for a long time (e.g. model 05). On the opposite end, models 02-04 had very high running times $\gtrsim 3 \times 10^5 GM/c^3$ and PNST.01 had an extreme high running time of $\sim 8 \times 10^5 GM/c^3$, which is the longest BH accretion flow simulation produced to date, to our knowledge\footnote{The previous longest-duration simulation is the three-dimensional GRMHD model of a RIAF performed by \cite{Chan2015}, which ran for $2.3\times 10^5 \ \grt$.}. 
\section{Results}\label{sec:results}

In this section, we present the results from the analysis of our numerical simulations. In subsections \ref{subsec:accretion-flow} and \ref{subsec:lag-part}, we present in detail the results for two of our models which illustrate the diversity of emergent behaviors, both in terms of initial conditions and the intensity of the resulting outflows: PNST.01 (very weak outflows) and PL2SS.3 (strong outflows). These two models are distinguished from the others because they reached inflow equilibrium up to large radii---in fact, they have the largest convergence radii among the models (cf. Appendix \ref{app:convergence}). In section \ref{subsec:integ} we present a holistic picture of the results from all our simulations.
In appendix \ref{subsec:other} we discuss the other simulations, which presented varying wind strenghts and convergence radii.

\subsection{Accretion flow properties}\label{subsec:accretion-flow}

\begin{figure*}
\center
\subfigure[][PNST.01]{\includegraphics[width=.7\linewidth]{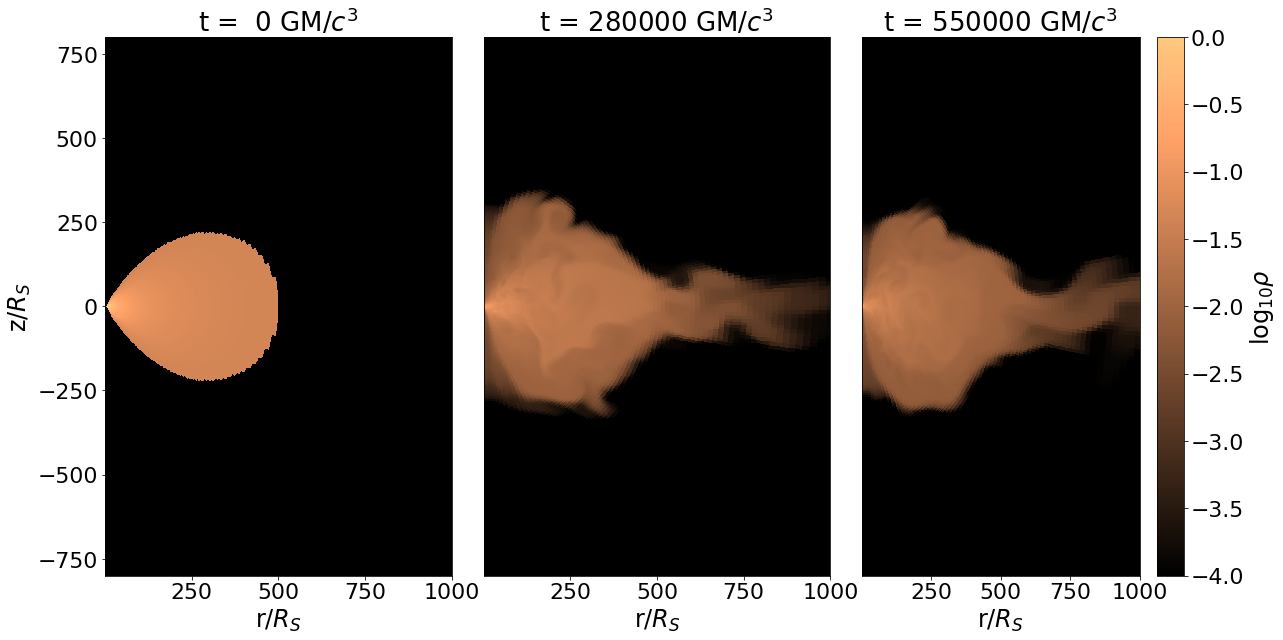}}
\qquad
\subfigure[][PL2SS.3]{\includegraphics[width=.7\linewidth]{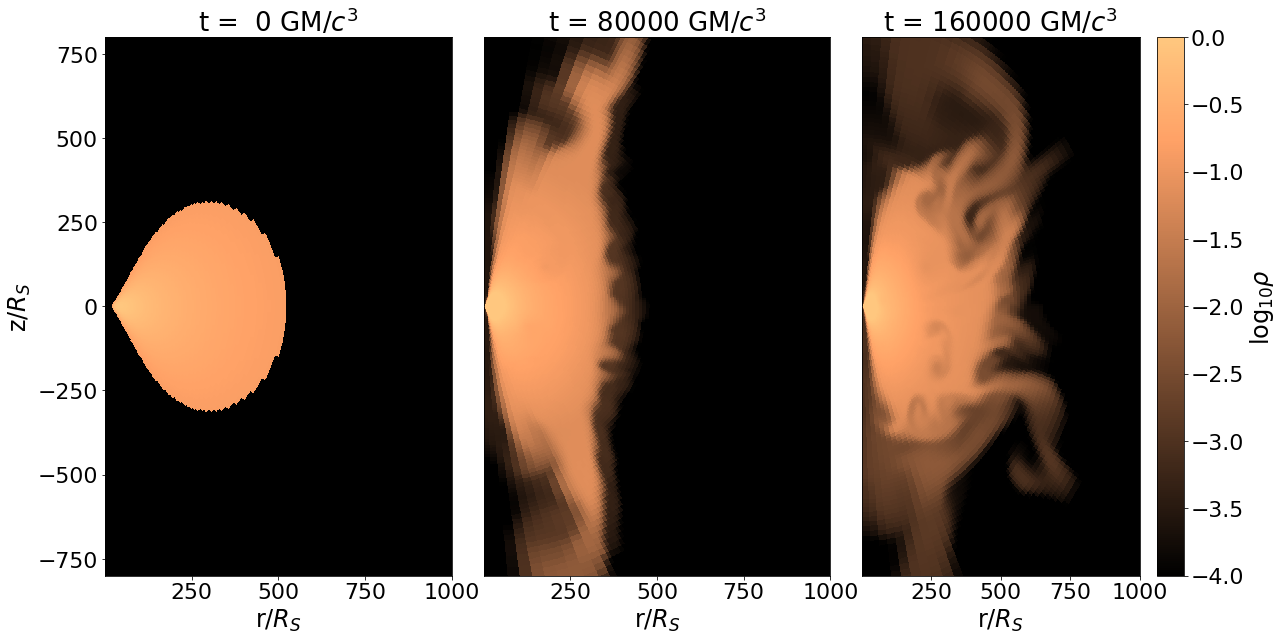}}
\caption{Snapshots of the density map for the main simulations where the color corresponds to $\log \rho(\mathbf{r})$. Here we can see how the torus evolves and changes its shape as time advances; in particular, we can see outflowing material reaching distances further than $500 R_S$.}
\label{fig:dens-maps}
\end{figure*}

Figure \ref{fig:dens-maps} shows snapshots of the density maps of models PNST.01 and PL2SS.3 at different times. Models PNST.01 presented a ``diffusion-shape'' and volume expansion of the torus, but not so dramatic as in model PL2SS.3. The bottom panel shows stronger ejection than the top one, with the formation of bipolar outflows and the torus shape becoming quite disturbed compared to its initial state. Model PL2SS.3--together with PL2SS.1--were the simulations that presented the strongest outflows. In the simulations above, we can see fluid elements being ejected to distances $\gtrsim 500R_S$--which is the initial torus equatorial outer edge adopted. 

\begin{figure*}
\center
\includegraphics[width=\linewidth]{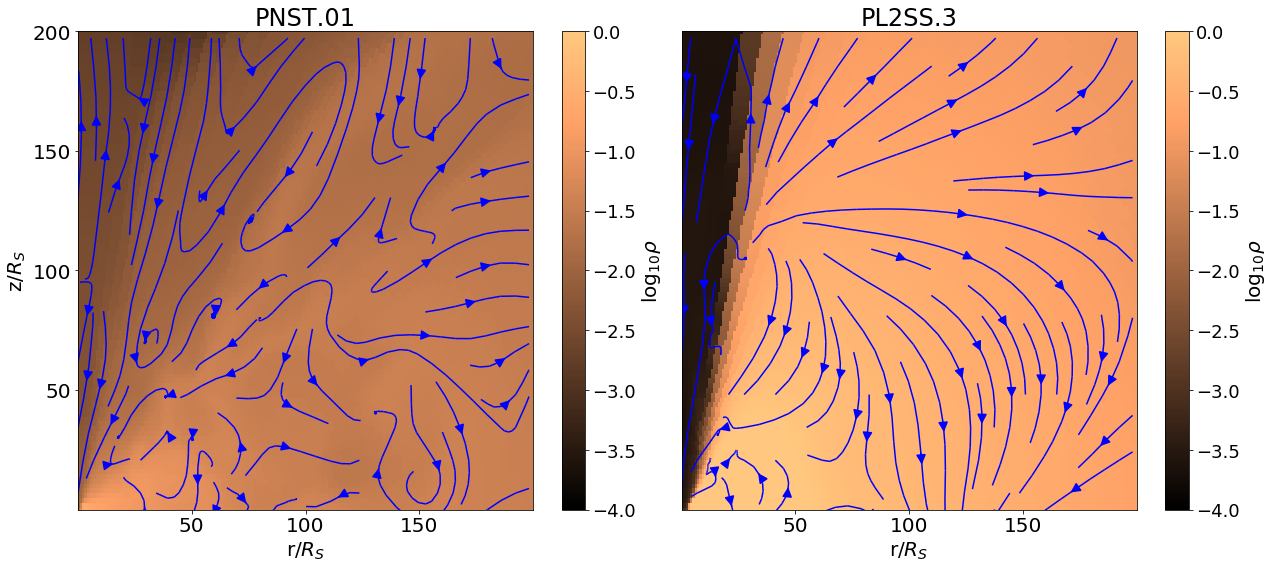}
\caption{Snapshots of the main simulations taken at $t \approx 160000 GM/c^3$. Here is the inner part of the accretion flow ($r < 200R_S$), the color corresponds to $\log \rho(\mathbf{r})$ and the blue arrows represent the velocity field.}
\label{fig:avg-dens}
\end{figure*}

\begin{figure*}
\center
\includegraphics[width=\linewidth]{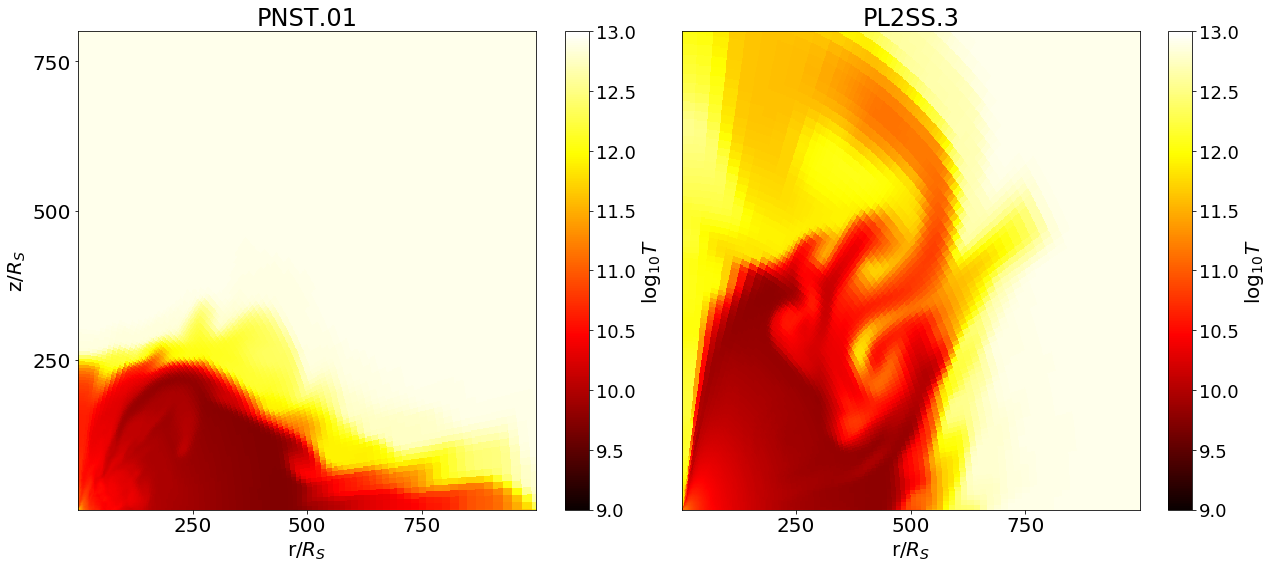}
\caption{Snapshots of the main simulations taken at $t \approx 160000 GM/c^3$. Here the color corresponds to $\log T(\mathbf{r})$. The white area corresponds to the low-density atmosphere around the initial torus. In these plots we observe the accretion disc surrounded by a hotter corona. The expelled material in PL2SS.3 is considerably hotter than the disc.}
\label{fig:avg-temp}
\end{figure*}

Figures \ref{fig:avg-dens} and \ref{fig:avg-temp} shows both the velocity field and the ion temperature distribution in our models. From the velocity field displayed, we can see that there is strong turbulence occurring in the accretion flow. From the bottom panel, we can see that the temperatures are quite high, as expected for RIAFs. The temperatures range between $10^{9}$ K near the equator to $\lesssim 10^{12}$ K towards the low-density regions in the corona and outflows. In the artificial atmosphere of the simulation (the white region in the plot) the temperature is even higher, reaching $10^{13}$ K, but this region have extremely low density and should not be taken into account in the analysis.

Following \cite{Stone1999}, we defined the accretion rate as the flux of material through a surface of radius $r$. We denoted $\dot{M}_{\rm in}$ the mass \textit{inflow} rate and $\dot{M}_{\rm out}$ the mass \textit{outflow} rate, which are defined as

\begin{align}
& \dot{M}_{\rm in}(r) = 2\pi r^2\int_0^{\pi} \rho \text{ min}(v_r, 0) \sin \theta d\theta, \label{mdot-in} \\
& \dot{M}_{\rm out}(r) = 2\pi r^2\int_0^{\pi} \rho \text{ max}(v_r, 0) \sin \theta d\theta.
\label{mdot-out}    
\end{align}

\begin{figure}
\noindent
\includegraphics[width=\linewidth]{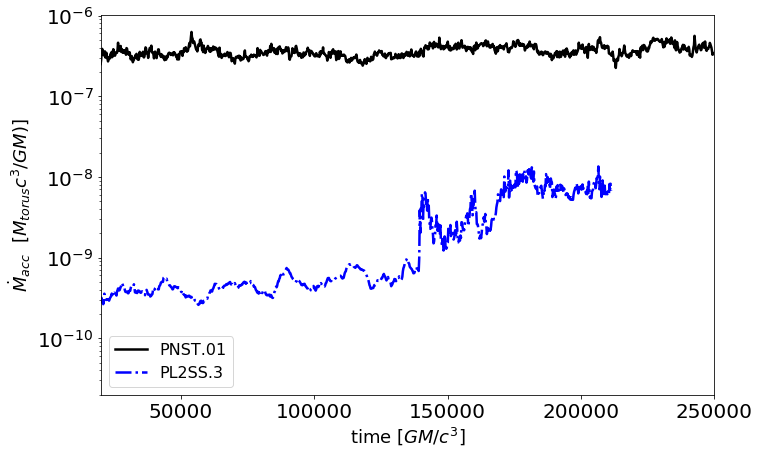}
\caption{Net mass accretion rate near the inner boundary of the simulation, $r = 1.5R_S$. Each line represents one of the three simulations, PNST.01 is the black solid line and PL2SS.3 is the dot-dashed blue line.}
\label{fig:acc-time}
\end{figure}

The net mass accretion rate is 

\begin{equation}
\dot{M}_{\rm acc} = \dot{M}_{\rm in} + \dot{M}_{\rm out}.
\end{equation}
Figure \ref{fig:acc-time} shows the net mass accretion rate calculated at the inner boundary of the simulation--which represents the event horizon\footnote{Note that since this is a Newtonian simulation, properly speaking we cannot define a perfectly absorbing event horizon boundary.}. Each line represents a different simulation. In this plot is very clear that the viscosity profile has strong impact in the mass accretion rate; for instance, simulations with the SS-viscosity have much weaker mass accretion rates. The accretion rates for PNST.01 and PL2SS.3 reach, respectively, a mean value of $10^{-6.5}$ and $10^{-(8.-9.)}$ in units of $M_0c^3/GM$ where $M_0$ is the torus initial total mass. 

\begin{figure}
\center
\subfigure[][PNST.01]{\includegraphics[width=\linewidth]{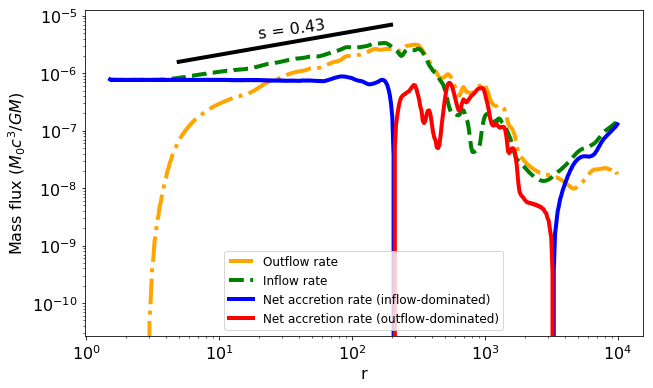}}
\qquad
\subfigure[][PL2SS.3]{\includegraphics[width=\linewidth]{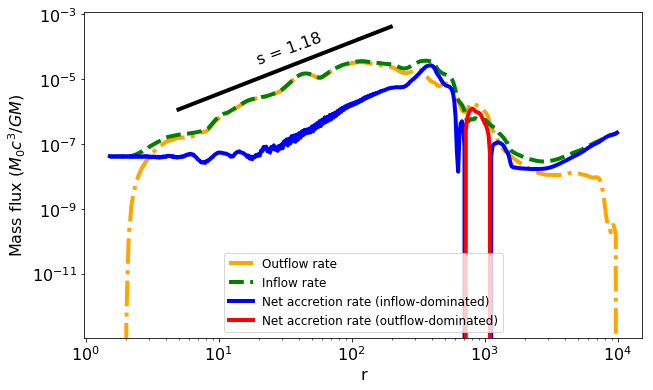}}
\caption{Mass flux radial profiles for the two main simulations, angle-averaged around the equatorial plane and time-averaged using the last 50 states of each model. The dash-dotted orange, dashed green and solid blue and red lines correspond to the inflow rate, outflow rate and net accretion rate, respectively. The color of the solid line indicates the dominant flow mode: blue if inflow dominates, red if outflow dominates. The solid black lines indicate the power-law fits to the inflow rates in the $20-200 R_S$ range --shifted upwards for clarity.}
\label{fig:acc-profiles}
\end{figure}

In figure \ref{fig:acc-profiles}, we show the radial dependence of the mass flux rates in the accretion flow; to obtain the mass flux here, we first computed the angle-average between $85\degree - 95 \degree$--i.e. around the equatorial plane--then we computed the time-average using the last 50 states of each simulation.
We find the most striking difference among the radial dependencies displayed in Figure \ref{fig:acc-profiles} is in the net accretion rates. For instance, in the ST model (top panel) we see a constant $\dot{M}_{\rm acc}$ until it starts to oscillate at a radius $200 R_S$. Conversely, in the SS simulations we have found a constant $\dot{M}_{\rm acc}$ until $r \sim 30R_S$, for $r \gtrsim 30R_S$ $\dot{M}_{\rm acc}$ increases until $\sim 500 R_S$ (model PL2SS.3, bottom panel). Furthermore, we see that the inflow rate in noticeably larger than the outflow rates, whereas in model PL2SS.3 the two curves closely track each other for most radii of interest. 

The inflow rates display a power-law radial dependence in the range $\approx 10-200 R_S$, in agreement with the \textit{ansatz} $\dot{M}_{\rm in} \propto r^s$ originally proposed by \cite{Blandford1999}. We fitted a $\dot{M}_{\rm in} \propto r^s$ curve to our simulation data in the radial range $20-200 R_S$ and the resulting fits are displayed in Figure \ref{fig:acc-profiles}. We find that $s$ ranges between $0.4$ and $2.6$--i.e. the power-law index of the dependence can be even higher than the value of one proposed by \cite{Begelman2012}.

\begin{figure}
\center
\subfigure[][PNST.01]{\includegraphics[width=\linewidth]{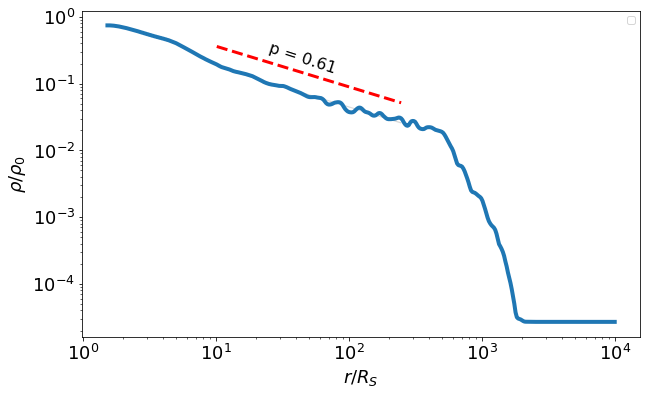}}
\qquad
\subfigure[][PL2SS.3]{\includegraphics[width=\linewidth]{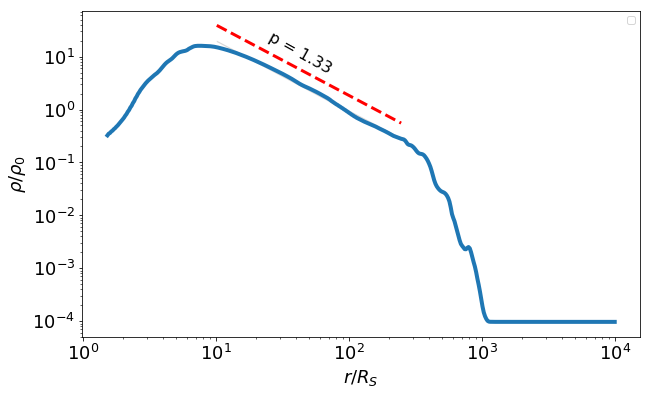}}
\caption{Density profiles for the two main simulations, $\rho(r)$, around the equatorial plane, it was angle averaged between $85\degree - 95 \degree$. These profiles were taken using the last 50 states of each model. The solid blue line is the density extracted from the simulation, the units are in code units of the defined $\rho_0$. The dashed red line is the adjust in the ``linear region'', adopted between $10-300R_S$.}
\label{fig:dens-profiles}
\end{figure}

The equatorial density profile in the accretion disc --computed in the same fashion as the mass flux described above-- is shown in Figure \ref{fig:dens-profiles}. As can be seen in the figure, the density is well-approximated by a power-law of the form $\rho \propto r^{-p}$ in the $r=10-300 R_S$ range, with the value of the power-law index $p$ in the range $0.6-1.5$ as indicated for each model in the panels. The resulting power-law dependence of $\rho(r)$ and the fact that $p < 1.5$ are in agreement with the general expectations of the ADIOS model \citep{Blandford1999}. It is also in agreement with previous hydrodynamical simulations \citep{Stone1999, Yuan2012b}. We compare our results with these models in section \ref{subsec:compare-sims}.

Finally, we provide a convenient conversion from $\dot{m}$ in code units to physical ones. The conversion is given by
\begin{equation}
\frac{\dot{M}}{\dot{M}_{\rm Edd}} = 3 \times 10^{-4} \left( \frac{M_{0}}{M_{\odot}} \right) \left( \frac{M_{\rm BH}}{10^{8} M_{\odot}} \right)^{-1} \left( \frac{\dot{M}_{\rm sim}}{\rm code \ units} \right),
\label{acc-edd}
\end{equation}
where $M_0$ is the initial torus mass, $M_{\rm BH}$ is the black hole mass and $\dot{M}_{\rm sim}$ is mass accretion rate in code units from the simulation. This is useful if one wants to read off e.g. the $\dot{m}$ variability values displayed in Fig. \ref{fig:acc-time} in physical units.

\subsection{Outflows and the Bernoulli parameter}

Traditionally, the Bernoulli parameter -- see equation \eqref{Be-eq} -- $Be$ has been used as an indicator of the presence of unbound gas in numerical simulations \citep{Narayan1994, Narayan2012, Yuan2012}. For a stationary, laminar flow, $Be$ can be interpreted as a quantity that measures how much the gas is gravitationally bound to the central mass. $Be < 0$ indicates a bound particle and $Be > 0$ a particle able to escape to infinity. This is the reason why positive values of $Be$ have been taken as indicating the presence of unbound outflows in numerical simulations of BH accretion. On the other hand, $Be > 0$ does not guarantee that a gas packet will be ejected, since $Be$ can change its sign in a viscous flow as discussed by \cite{Yuan2015}. In any case, we analyzed the behavior of $Be$ in our models. In our simulations, $Be$ is positive in most parts of the flow with the exception of the innermost parts located at $r \lesssim 50 R_S$. 

\subsection{Efficiency of wind production}  \label{subsec:eff}

We now present our results related to the energetics of the winds produced in our simulations. Quantifying the energy outflows from SMBHs is instrumental in the understanding of the coevolution between SMBHs and their host galaxies, since the energy deposited by BH winds can potentially offset gas cooling and quench star formation (cf. introduction). From our simulations, we are able to compute separately the energy outflow rate through winds, $\dot{E}_{\rm wind}$, and the mass accretion rate onto the BH, $\dot{M}$. We then defined a ``wind efficiency factor'' $\eta$ as

\begin{equation}
\dot{E}_{\rm wind} = \eta \dot{M} c^2.
\label{eta-efficiency}
\end{equation}
which is the quantity we quote in this paper. Before turning to this efficiency, we need to define what we mean by $\dot{E}_{\rm wind}$ and $\dot{M}$.

Typically, in applications of AGN feedback such as cosmological simulations of galaxy evolution, the authors estimate the feedback power from a mass accretion rate provided to the BH near its Bondi radius $R_{\rm Bondi}$--usually the Bondi accretion rate (e.g. \citealt{Di2005, Sijacki2015}). For consistency with such works, in our simulations we defined $\dot{M}$ in equation \ref{eta-efficiency} as the mass accretion rate at the initial outer radius $R_{\rm out}$ of our accretion flow,

\begin{equation}
\dot{M} \equiv \dot{M}_{\rm in} (R_{\rm out})
\end{equation}
which is computed using equation \ref{mdot-in}. We choose to compute $\dot{M}$ at this radius because in our case this is a more appropriate estimate of the outer accretion rate.

The energy outflow rate was calculated as the surface integral

\begin{equation}
\dot{E}_{\rm wind} = \int \epsilon \max (v_r,0) dA 
\label{ewind-definition}
\end{equation}
calculated at $r=R_{\rm out}$ and only within the angle intervals $15\degree \leq \theta \leq 45\degree$ or $135\degree \leq \theta \leq 165\degree$ as defined in section \ref{subsec:lag-part}. With the integral defined in the above equation, when computing the energy rate we will automatically consider only fluid elements with $v_r > 0$. $\epsilon$ is the energy density taking into account the kinetic, thermal and gravitational contributions, defined as

\begin{equation}
\epsilon (\textbf{r}) = \rho(\textbf{r}) \frac{v(\textbf{r})^2}{2} + \frac{\gamma}{\gamma -1}p(\textbf{r}) - \frac{GM\rho(\textbf{r})}{R- R_S}.
\label{specific-energy}
\end{equation}
Therefore, $\dot{E}_{wind}$ is the total power (minus rest mass energy) carried by outflowing gas that crosses the spherical surface at $R=R_{\rm out}$, not taking into account the poles and the accretion disc domain.

\begin{figure}
\noindent
\includegraphics[width=\linewidth]{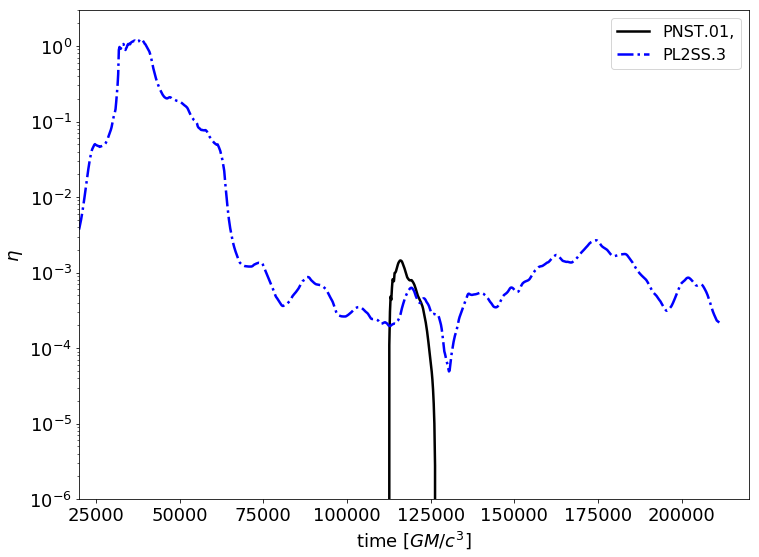}
\caption{Temporal evolution of the wind efficiency $\eta$ as defined in equation \ref{eta-efficiency} for the simulations PNST.01 (solid black line) and PL2SS.3 (dot-dashed blue line).}
\label{fig:eta-winds}
\end{figure}

Now we are in a position to present the resulting efficiency of wind production. The temporal evolution of $\eta$ for the two main simulations is presented in Figure \ref{fig:eta-winds}. Each simulation had a strikingly different behavior of $\eta(t)$ with respect to each other. The strongest winds are found in model PL2SS.3--supporting the conclusion from the  density maps in Figure \ref{fig:dens-maps}. For instance, at $t \sim 50000 GM/c^3$ the efficiency peaks at $\eta \approx 1$, i.e. the wind power is comparable to the instantaneous accretion power. Afterwards, $\eta$ drops to a flat value around $10^{-3}$ in the remaining simulation time. For model PNST.01 there is no continuous outflow. Instead, model PNST.01 displays only a timid outflow burst at $t \sim 1.2 \times 10^5 GM/c^3$ with a peak of $\eta \approx 10^{-3}$, lasting for $\Delta t \approx 1 \times 10^4 \grt$. Despite $\eta$'s variability in all models, we did not find any clear periodic oscillation.

\subsection{Analysis using tracer particles}\label{subsec:lag-part}
One of the strengths of using the technique of tracer particles (section \ref{subsec:traj-app}) is that we are able to quantify more precisely the amount of mass lost from the disc due to outflows by tracking the amount of mass carried by each particle. 

\begin{figure}
\noindent
\includegraphics[width=\linewidth]{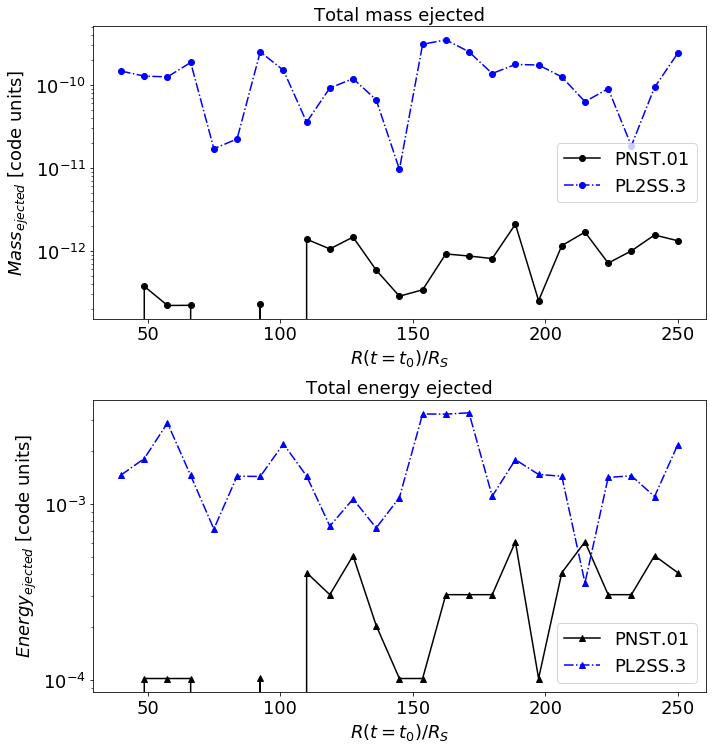}
\caption{Total ejected mass and energy in the main simulations in arbitrary units. Models PNST.01 and PL2SS.3 are displayed as solid black line and dot-dashed blue line, respectively. The x-axis is the launching radius of the particle, and y-axis is particle's final mass/energy. We can see that the loss of both mass and energy is more pronounced in model PL2SS.3 compared to the other one--i.e. the resulting outflows in this model are stronger.}
\label{fig:LP-mass-energy}
\end{figure}

In Figure \ref{fig:LP-mass-energy} we show the mass and energy carried away by the outflowing particles following the criteria defined in section \ref{subsec:traj-app} for the ``real outflow''. We defined the relative fraction of ejected mass $f_{\rm m}$ and the fraction of ejected energy $f_{\rm e}$ as 

\begin{align}
f_{\rm m} &     = \frac{\rm mass\ in\ tracer\ particles\ lost\ in\ outflows}{\rm total\ mass\ of\ tracer\ particles} \nonumber \\ 
& = \frac{\sum_k \rho_k(t=t_{\rm final}, r = r_{\rm final}) \delta V \Theta[r_{k}(t=t_{\rm final}) - r_{\rm out}]}{\sum_k \rho_k(t=t_{\rm initial}, r = r_{\rm initial}) \delta V}, \label{mass-ejected-real} 
\end{align}
where the sums are carried over all tracer particles and $\Theta$ is the Heaviside function. The mass of each particle was defined as: 
\begin{equation}
m_k (t) =  \rho[\textbf{r}_k(t)] \delta V,
\end{equation}
where we assume that all particles occupy the same small volume $\delta V = \rm const$. The specific value that we adopt for $\delta V$ does not matter because when computing $f_{\rm m}$ using equation \ref{mass-ejected-real}, $\delta V$ cancels out. Similarly, we defined the relative fraction of ejected energy $f_{\rm e}$ as
\begin{equation}    
\label{energy-ejected-real}
f_{\rm e} = \frac{\sum_k E_k(t=t_{\rm final})\Theta[r_{k}(t=t_{\rm final}) - r_{\rm out}]}{\sum_k E_k(t=t_0)}
\end{equation}
where the energy is defined as $E(\textbf{r}) = \epsilon(\textbf{r}) \delta V$  and $\epsilon$ is the energy density from \eqref{specific-energy}. 

The mass ejection plot is the upper one in the panels of figure  \ref{fig:LP-mass-energy}, whereas the energy ejection is displayed in bottom panel. From these two plots we can see that the ejected energy roughly follows the same pattern as the mass ejection.

In addition the behavior of mass (or energy) loss are similar at all radii. Considering the original mass (or energy)--see equations \eqref{mass-ejected-real} and \eqref{energy-ejected-real}--, the simulations presented a fraction of mass (energy) ejected of up to 0.2\% (2\%) of the total mass (energy) available, with an average value of 0.03\% (0.2\%). The difference between these two values can be attributed to the different temperatures in the accretion disc and the corona. The particles were heated and accelerated away as the disc's corona thermally expands, carrying out energy.

\begin{figure*}
\noindent
\makebox[\textwidth]{
\subfigure[][PNST.01 ]{\includegraphics[width=6.5cm]{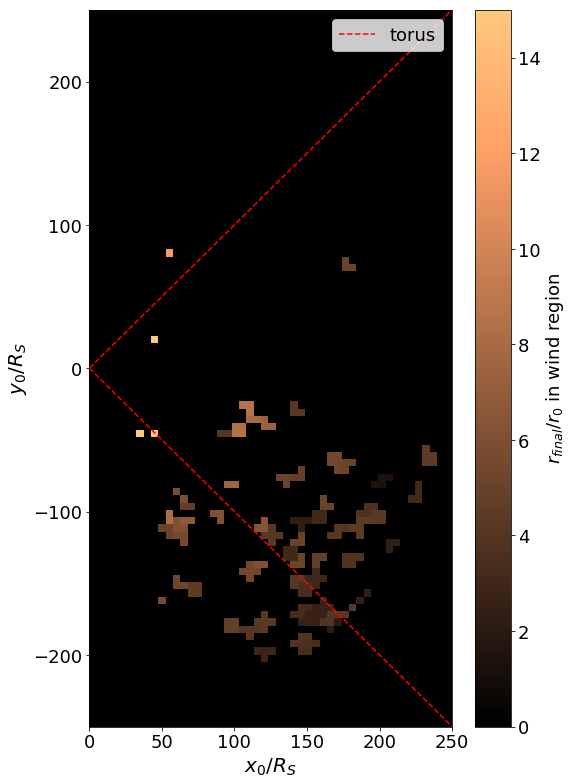}}
\subfigure[][PL2SS.3]{\includegraphics[width=6.5cm]{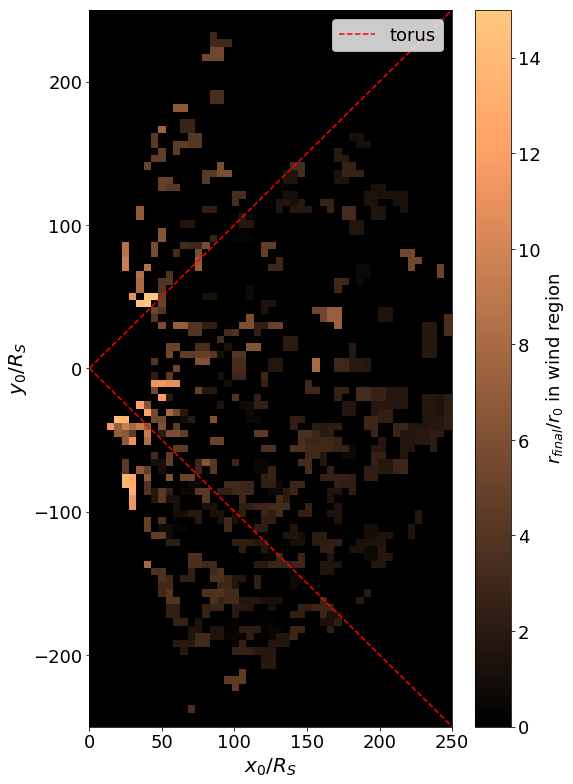}}
}
\caption{Maps of the wind depth illustrating the regions of the accretion flow from which outflows are produced. Lighter regions eject particles which reach farther distances compared to the darker regions.}
\label{fig:LP-map}
\end{figure*}

Mass-loss through winds is not uniformly distributed across all radii. In order to quantify how far a particle originated in a certain radius can go, we plotted the quantity $r(t_{\rm final})/r(t_0)$--which we will refer to as wind depth henceforth-- in Figure \ref{fig:LP-map}. Larger values of the wind depth in a given region of the flow indicate that it can produce outflows that reach large distances. As such, Figure \ref{fig:LP-map} is tracking the accretion flow regions where the ejected particles come from. The two panels were labeled for each simulation and we considered only particles that are in the wind region. In model PL2SS.3 we see bipolar outflows, whereas model PNST.01 displays a strange asymmetry --a unipolar outflow-- with all the ejections occurring in the same side, which is very unique when compared with the other simulations we performed. This behavior is qualitatively similar to the unipolar outflows seen in model G of \cite{Igumenshchev2000} (cf. Fig. 12 in that paper). In model PNST.01, the ejection occurred mainly in the torus corona -- similarly to coronally-driven winds--whereas model PLSS.3 seems to produce winds from all regions of the disc with a more homogeneous ejection region, with outflows coming even from close to the equator. 

\begin{figure}
\noindent
\subfigure[][PNST.01]{\includegraphics[width=\linewidth]{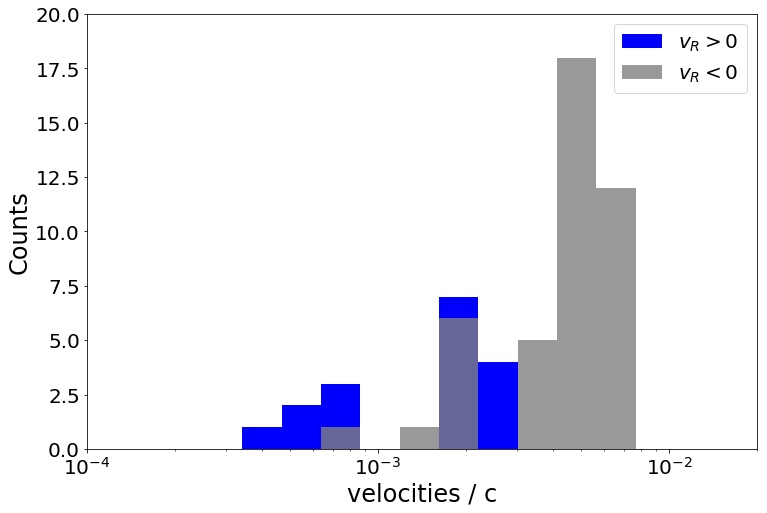}}
\qquad
\subfigure[][PL2SS.3]{\includegraphics[width=\linewidth]{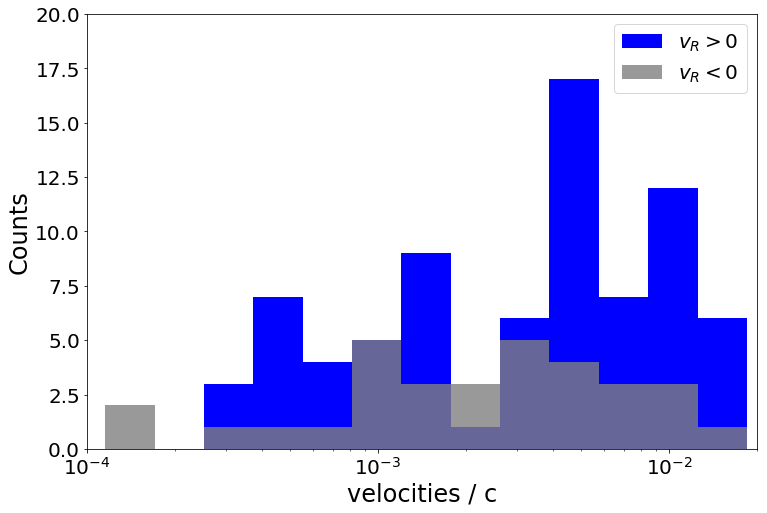}}
\caption{Distribution of velocities of the ejected particles for simulations PNST.01 and PLSS.3. These histograms displays the averaged velocity of the ejected particles in the last $\sim 1000 GM/c^3$ of each simulation. The blue columns represented the population of particles with $v_r > 0$ (outflow), the grey columns represented the population of particles with $v_r < 0$ (fallback).}
\label{fig:LP-velocities}
\end{figure}

An important parameter to be analyzed in these simulations are the velocity of these ejected particles. The distribution of their velocities is displayed in Figure \ref{fig:LP-velocities}. In the figure we divided the sample in two types of particles, the ones with $v_r > 0$ (blue) --which we refer to as ``outflow'' particles since they are expected to be in outflows-- and the other ones with $v_r < 0$ (grey) which fall back and can be reincorporated into the accretion disc --the latter types of particles are referred to as ``fallback''. PNST.01 had a low rate of outflow particles and is dominated by fallback ones which reach the highest velocities of the simulation. PL2SS.3 was dominated by outflow particles. Considering only the outflow particles, the average velocities for outflow particles for simulations PNST.01 and PL2SS.3 were respectively $1.6 \times 10^{-3}c$ and $5.0 \times 10^{-3}c$. For all simulations $\overline{v}_{\rm out}$ was in the range 0.001-0.005c. The ejected particles presented nonrelativistic velocities. For instance, the maximum velocity of an individual particle in the simulations did not exceed $0.05c$. All these ejected particles --both ``outflow'' and ``fallback''-- had a positive value for the Be.

\subsection{Overview of results for all models} \label{subsec:integ}

\begin{figure}
\noindent
\includegraphics[width=8.0cm]{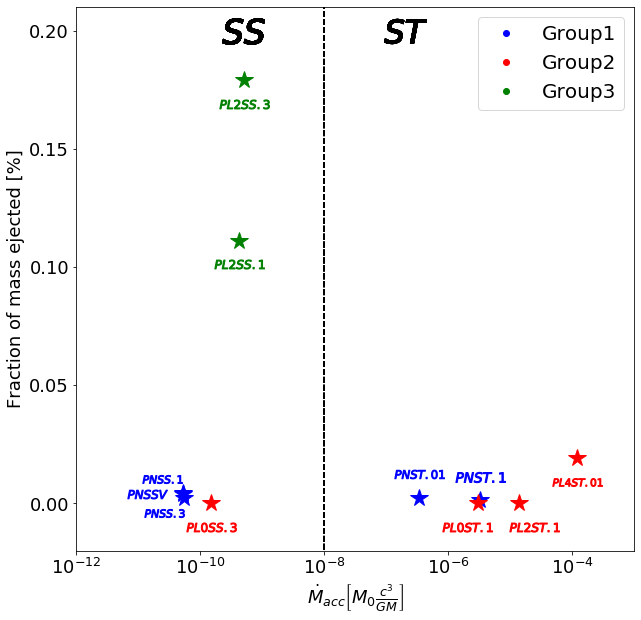}
\caption{Net mass accretion rate $\dot{m}$ versus the fraction of ejected mass of the simulations. The labels identify the simulations. We divided them in three groups for the analysis as described in the text. The black dotted line in the center are separating the two regime of viscosity adopted, in the left-side there is the simulations with SS-viscosity, in the right side the ones with ST-viscosity (see section \ref{subsec:equations}).}
\label{fig:integrated-mdot}
\end{figure}

After the individual analysis of each simulation we proceed to analyze these results as a whole. Table \ref{tab:allmodels} shows the results for all simulations that we computed. In Figure \ref{fig:integrated-mdot} we plotted $f_{\rm m}$ as a function of $\dot{m}(1.25R_S)$,  i.e. it relates our the fraction of mass  lost in the wind (cf. equation  \ref{mass-ejected-real}) and the net mass accretion rate at the event horizon (more rigorously, at the inner boundary of the simulation). $\dot{m}$ is normalized by the torus initial mass assuming that all simulations had the same total torus mass in the beginning. 

\begin{table*}
\noindent
\centering
\begin{threeparttable}
\begin{tabular}{lcccccccccccc}
\hline \hline 
\multicolumn{1}{c}{Short} &
\multicolumn{1}{c}{Full} &
\multicolumn{1}{c}{$p^{\rm 3}$} &
\multicolumn{1}{c}{$s^{\rm 4}$} &
\multicolumn{1}{c}{$\eta^{\rm 5}$} &
\multicolumn{1}{c}{$\max(\eta)$} &
\multicolumn{1}{c}{Wind} &
\multicolumn{1}{c}{$f_{\rm m}$$^{\rm 7}$} &
\multicolumn{1}{c}{$f_{\rm e}$$^{\rm 8}$} &
\multicolumn{1}{c}{$\overline{v}$$^{\rm 9}$} &
\\
name$^{\rm 1}$ & name$^{\rm 2}$ &  &  & ($\times 10^{-3}$) &   & activity           & (\%) & (\%) &  ($c$) \\
               &                &  &  & &  & time$^{\rm 6}$ (\%)&      &      &         &       \\
\hline \hline 
00 & PNST.01 & $0.61 $ & $0.43 \pm 0.01$ & 0.0 & 0.014  & 2  & 0.002 & 0.55 & 0.0020 \\
01 & PNST.1  &$0.89 $ & $0.17 \pm 0.01$ & 0.0  & 0.060  & 15 & 0.001 & 0.23 & 0.0062 \\
02 & PNSS.1  &$1.16 $ & $2.55 \pm 0.03$ & 0.2  & 0.22 & 51 & 0.004 & 0.92 & 0.0010 \\
03 & PNSS.3  &$1.16 $ & $2.19 \pm 0.04$ & 0.0  & 0.21 & 46 & 0.002 & 0.53 & 0.0010 \\
04 & PNSSV   &$1.16 $ & $2.61 \pm 0.02$ & 1.2  & 0.22 & 53 & 0.0 & 0.79 & 0.0018 \\
05 & PL0ST.1 &$0.97 $ & $-0.11 \pm 0.01$ &0.0  & 0.53 & 13 & 0.0 & 0.0 & -- \\
06 & PL0SS.3 &$0.91 $ & $1.08 \pm 0.04$ & 15$^{10}$&7.6& 98 & 0.0 & 0.0 & -- \\
07 & PL2SS.1 &$1.37 $ & $0.77 \pm 0.05$ & 6.5  & 12& 95 & 0.11 & 0.60 & 0.0028 \\
08 & PL2SS.3 &$1.33 $ & $1.18 \pm 0.04$ & 7.4  & 12& 97 & 0.18 & 0.31 & 0.0045 \\
09 & PL2ST.1 &$1.13 $ & $0.02 \pm 0.01$ & 0.0  & 33& 45 & 0.0 & 0.0 & -- \\
10 & PL4ST.01&$1.53 $ & $0.10 \pm 0.01$ & 7.4  & 10& 56 & 0.019 & 1.77 & 0.0017 \\
\hline \hline \\
\end{tabular}
\begin{tablenotes}\footnotesize
\item[1] Short model name.
\item[2] Full model name including information on parameters.
\item[3] Power-law coefficient defined as $\rho \propto r^{-p}$. The $1\sigma$ uncertainty corresponds to $0.01$ from the fits. 
\item[4] Power-law coefficient defined as $\dot{M}_{\rm in} \propto r^{s}$.
\item[5] Median value of $\eta(t) \times 10^{-3}$
\item[6] Fraction of the total time in which $\eta > 0$.
\item[7] Fraction of the mass ejected following the lagrangian particle analysis (see \eqref{energy-ejected-real}).
\item[8] Fraction of the energy ejected following the lagrangian particle analysis (see \eqref{energy-ejected-real}).
\item[9] Refers to Lagrangian particles.
\item[10] Unusually high value, better discussed in section \ref{app:PL0SS3}
\end{tablenotes}
\end{threeparttable}
\label{tab:allmodels}
\caption{Results concerning outflows for all simulations.}
\end{table*}

Each simulation occupies a different region of the diagram in Figure \ref{fig:integrated-mdot}. The different viscosity parameterizations adopted are clearly distinguishable, for instance simulations with the ST prescription generated $\dot{m}$ values orders of magnitude higher than the SS profile. Motivated by this considerable difference, we plotted the black dotted line in the figure to separates these two types of simulations. We divided them in three groups for the analysis:
\begin{itemize}
\item Group 1: simulations with the specific angular momentum adapted from \cite{Penna2013};
\item Group 2: simulations with power-law $l(R)$ and smallest fraction of ejected mass;
\item Group 3: simulations with power-law $l(R)$ and highest fraction of ejected mass.
\end{itemize}
They have some major characteristics considering both fluid and particle analysis:
\begin{itemize}
    \item Group 1 had on average $.01\%$ of mass ejection, this value seems that does not change drastically with the free parameters of the simulation or the adopted viscosity. The wind flux (see equations \eqref{ewind-definition}-\eqref{eta-efficiency} and figure \ref{fig:eta-winds}) of these simulations was non-continuous, winds were not generated all the time here. The simulations with SS viscosity presented a very small convergence radius. The average velocity of the ejected particles here are smaller than the group averaged velocity for Group 3, $\overline{v}_{\rm out}^{\text{    G1}} \lesssim \overline{v}_{\rm out}^{\text{    G3}}$. 
    \item Group 2 had the smallest fraction of mass ejected, except for PL4ST.01. These simulations presented strong inflow component, except for PL0SS.3, the inflow was so intense in these three that suppressed any outflow. PL0SS.3 did not present the same inflow component as the other ones, but the particles remained inside the initial torus all the way (see first panel from figure \ref{fig:initial-conditions}). The wind generation pattern of these simulations varied for all simulations. This group presented completely heterogeneous properties.
    \item Group 3 are the simulations with the most energetic winds and particles. Models PL2SS.1 and PL2SS.3 are very similar simulations with the only difference in the value of $\alpha$, as discussed before. The setup consisting of $a = 0.2$ and SS-viscosity presented powerful outflows, with a continuous generation of winds, and some of highest average velocities from our sample $v \approx 0.003-0.004$c.
\end{itemize}

\begin{figure}
\noindent
\includegraphics[width=8.0cm]{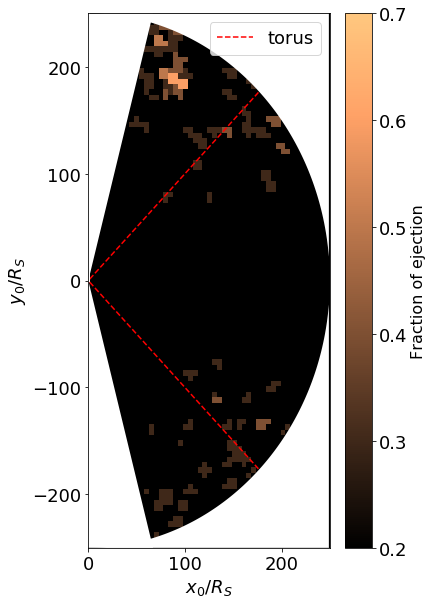}
\caption{Map showing the average fraction of particles in a given position which are lost in outflows, taken over all simulations carried out in this work (variable $\langle \Xi \rangle$, equation \ref{ej-map}).  }
\label{fig:integrated-map}
\end{figure}

It is worthwhile asking: considering holistically all the models which produced winds, what is the location in the disc from which the outflowing particles come from, on average? For this purpose, we apply the tracer particles formalism to locate the launching region in the eleven simulations. For each model, we considered only the particle ejected in the wind region -- similarly to Figure \ref{fig:LP-map} -- by defining the binary variable
\begin{equation}
    \begin{aligned}
\Xi = 
\begin{cases}
     1,            & \text{if }  (\textbf{r}_{\rm final} \text{ is in wind region) and} \ (r_{\rm final} > R(t_0)) \\
     0,  & \text{otherwise}.
\end{cases}
\end{aligned}
\label{ej-map}
\end{equation}
The variable $\Xi$ informs whether a particle at a given position has been ejected ($\Xi=1$) or not ($\Xi=0$). After creating maps of $\Xi$ for all simulations, we added them up and computed the average, $\langle \Xi \rangle$. The result can be seen in figure \ref{fig:integrated-map}, where the color scale indicates the likelihood that a particle  located at the given position at the beginning of all simulations becomes part of an outflow later on. A value of one at a certain position would indicate that in all simulations a particle initially at that position was ejected; conversely, a value of zero means that in all simulations a particle initially at that position was not ejected. We can see in Figure \ref{fig:integrated-map} the presence of some regions with values of ejected particles in $\sim 50 \%$ of the simulations (i.e. with values $\langle \Xi \rangle >0.5$). These regions with higher likelihoods of producing winds are located in the corona of the accretion disc, suggesting that the winds we are seeing correspond to coronal winds. 

Finally, we computed the power spectrum from the time series of different quantities such as $\eta$ and mass accretion rate. We did not find periodicity in any of the simulations. It is possible that our 2D setup suppressed any possible orbital related variability, since we assumed axial symmetry.
\section{Discussion}\label{sec:disc}

\subsection{Accretion flow and density radial profile}

In table \ref{tab:allmodels} we present the power-law index $p$ for density radial profile $\rho \propto r^{-p}$ averaged over the equatorial region of the accretion flow. From this table we can draw a number of conclusions:
\begin{enumerate}
\item There is correlation between the initial angular momentum profile adopted and the value of $p$. The corollary is that we see no particular values of $p$ associated with any of the three groups in figure \ref{fig:integrated-mdot}. 
\item For simulations with the same specific angular momentum, the SS-viscosity models resulted in higher values of $p$ compared to the ST-viscosity ones.
\item Our results did not present clear correlation between the value of $p$ and the wind production or $s$.
\end{enumerate}

The last item above is especially relevant because it demonstrates that based only on the value of $p$, it is not straightforward to tell whether there are winds being produced. This result seems to contradict some previous analytical \citep{Blandford1999, Begelman2012} and numerical \citep{Yuan2012} works which base their analysis on the assumption that $\rho(r)$ in the accretion disc is strongly dependent on the presence of mass-loss. These works assume that $\rho(r) \propto r^{-3/2+s}$ where $s$ is usually in the range 0.5-1 with larger values corresponding to more profuse outflows ($s=0$ corresponds to a no-wind ADAF; \citealt{Narayan1994}). Concretely, ADIOS models suggest that $s=1$, $p=0.5$ corresponds to very strong winds. Our model PNST.01 shows such a similar density profile, however it display a feeble breeze over just a short amount of time. Our model with the strongest winds--model PL2SS.3--has a low value of $p=1.33$ in contradiction with ADIOS models, and also similar to models with no winds such as PL2ST.1. We conclude that we cannot make strong statements about the presence of winds based on the indirect information given by $\rho(r)$. 

$s$ was clearly related to the adopted viscosity for the simulation, with ST simulations showing $s \lesssim 0.5$, while SS simulation had much higher values as $0.75 < s < 1.2$ --some simulations reached $s > 2$, but they presented poor convergence, see table \ref{tab:convergence}. Furthermore the relation between $p$ and $s$ are not clear in our simulations sample, despite the expected relation in ADIOS models of $s + p = 3/2$. The values of these two power-law index are more related to the viscosity and initial conditions than to each other. In fact, they are probably non-trivial functions of the flow parameters.

The role of viscosity is prominent in our results and the evolution of accretion rate is directly related to it. The SS-viscosity produced simulations with lower accretion rates, as we can see in figure \ref{fig:integrated-mdot}. $\dot{m}$ is a direct consequence of the angular momentum transfer. ST-viscosity simulations transferred angular momentum more efficiently than SS-viscosity simulations. For SS-viscosity case the accretion is slower and we have steeper density and accretion rate profiles --respectively the values of $p$ ans $s$. For higher accretion rates, the gas was not able to accumulate in the inner regions of the accretion flow, it was accreted. For instance, for ST-viscosity we have $\max(\rho) \approx 1$ while for SS-viscosity we have $\max(\rho) > 1$.

The wind production is more persistent for SS-viscosity simulations. Our results suggested that lower accretion rates are more prone to produce outflows. Lower accretion means that the material have slower radial velocity and remains more time trapped inside the accretion flow. These effects make the material more likely to be subject to internal turbulent forces and thermal expansion.

Group 3 in figure \ref{fig:integrated-mdot} (the green ones) are the simulations with higher mass loss and a more steady outflow production. These two simulations had the same SS-viscosity profile and specific angular momentum profile $l(R) \propto r^{0.2}$. The initial disc can be seen in figure \ref{fig:initial-conditions}, panel (b). Higher values of $a$ means higher axial velocities and higher values of energy. These systems have naturally higher $Be$ associated, make them more prone to produce outflows.

The third of our explored parameters is $\alpha$. The effects of $\alpha$ are not clear as the others. Simulations PL2SS.1 and PL2SS.3 were very similar, with a major difference only in the values of $s$. Otherwise simulations PNST.01 and PNST.1 presented notable differences. $\alpha$ is related to the ``strenght'' of the viscous effects. For the same viscosity prescription and specific angular momentum profile, the simulation with higher $\alpha$ presented higher $\dot{m}$ for the pairs cited. 

\subsection{Wind launching mechanism}

Since our simulations do not have magnetic fields that could be responsible for ejecting material though the Lorentz force, the only possibility left is a thermally-driven mechanism to explain our observed outflows. In order to interpret the hydrodynamic winds observed here, we use the model of \cite{Parker1960} originally proposed to explain the nature of the Sun's coronal outflows. The main parameter that describes Parker winds is the ratio between gravitational binding energy and thermal energy

\begin{equation}
\Lambda = \frac{2 G M m_H}{5 r k T(r)}
\end{equation}
where $m_H$ is the hydrogen mass and $k$ is the Boltzmann constant. 
For $\Lambda \leq 1$, the thermal energy overcomes the gravitational energy and winds can be thermally launched via thermal expansion.

\cite{Parker1960} originally considered spherically symmetric mass outflows in stellar system with temperatures $\sim 10^6$K which are much lower than the typical temperatures in RIAFs. The question of course is: can much hotter accretion flows launch thermally-driven (Parker) winds, even though the central mass is quite larger than in stellar systems? \cite{Waters2012} attacked this question in the context of much colder, thin accretion discs around BHs. Here, we analyzed it in the context of our hot accretion flow models. More recently \cite{Cui2019} worked with solutions of Parker winds from RIAFs around SMBHs.

Analyzing the averaged temperature profile of our simulations we found $\Lambda \sim 1-2$ in the disc equator and $\Lambda \ll 1$ in the coronal region. Therefore, the winds we have observed in our RIAF models are consistent with being launched from the RIAF's corona via Parker wind scenario. Our outflows are produced by the extremely high temperatures in the torus corona. In this region we have $\Lambda \ll 1$ favouring thermal expansion and material ejection via thermally-driven winds.

PNST.01 shows only a short wind burst while PL2SS.3 displays persistent, vigorous mass-loss over the entire simulation time. These result come from the interplay between thermal expansion (i.e. Parker winds), the initial reservoir of angular momentum and shear stress. The difference between these models lies in the parameters which regulate the angular momentum and its transport across the flow. The initial launching of winds in both models is similar since they have similar amounts of enthalpy compared to the gas gravitational binding energy---both simulations had similar initial profiles of Bernoulli parameter and $\Lambda$. However, model PL2SS.3 has initially $\sim 33\%$ more angular momentum than the PNST.01 setup. The wind in PL2SS.3 carried a considerable amount of angular momentum away from the mid-plane of the accretion flow. While in PNST.01 there is no persistent wind and the transport of angular momentum outwards occurred in the equatorial zone, generating a tail-like structure.

With our very long simulations, we have found that the wind production is not continuous in time as can be seen in figure \ref{fig:eta-winds} --for PNST.01-- and in table \ref{tab:allmodels}. For Parker winds, with $\Lambda \leq 1$ we can reach, or not, stationary expansion solutions (continuously outflow generation), however the coronal heating can be not sufficient to cause the stationary expansion state even with the $\Lambda$ condition achieved, in this case found an intermittent expansion state \citep{Parker1960}, which match with our results. In $7^{th}$ column of table \ref{tab:allmodels} there is the fraction of time in each simulation in which $\eta > 0$, in other words the fraction of time in which the system ejected material.

We have found that a small change in the value of the $\alpha$-viscosity can have a notable effect on the properties of the resulting outflow. For instance, consider the models PL2SS.1 and PL2SS.3. A small increase in the value of $\alpha$ from 0.1 to 0.3 resulted in a notable decrease in the amount of energy carried by the outflow as we can see in the $9^{\rm th}$ column in table \ref{tab:allmodels}. Interestingly, the accretion rate did not change with this variation. A possible qualitative explanation is that for small values of $\alpha$ there is not enough gas reaching the wind launching region, so the wind is very weak or absent. On the other hand, with very high values of $\alpha$ there is enough gas being channeled in an outflow but the increased viscosity makes it lose energy and angular momentum rapidly. Therefore, there would an intermediate ``sweet spot'' of $\alpha$-values that optimizes wind launching, such that enough gas is lost in an outflow and keeping it stable and with enough energy to reach large distances. 

We found that the SS viscosity profile is the most conducive to wind formation. Simulations with the specific angular momentum scaled as a power-law ($r^a$) with higher values for the coefficient $a$ presented stronger winds. Simulations PL2SS.1 and PL2SS.3 were the ones with the most prominent outflows. Changing values of $\alpha$ did not change drastically the wind production, this is visible when we compare similar simulations such as PL2SS.1 and PL2SS.3.

\subsection{Comparison with observations}\label{subsec:obs}

Our simulations with the ST viscosity (models PNST.01, PNST.1, PL0SST.1 and PL2ST.1) resulted in values $p \sim 0.5-1$. The resulting density profiles are consistent with those constrained from observations of LLAGNs, for instance Sgr A* ($p \sim 0.5$; \citealt{Yuan2003, Wang2013}), NGC 3115 ($p \sim 1$; \citealt{Wong2011, Wong2014, Almeida2018}) and M87 ($p \sim 1$; \citealt{Kuo2014, Russell2015, Park2019}). In our sample these simulations had weaker winds compared with the remaining ones. The simulations with SS viscosity (models PNSS.1, PNSS.3, PNSSV, PL0SS.3, PL2SS.1, PL2SS.3) achieved more efficient winds but with $p \sim 1.1-1.4$, marginally consistent with the observations of NGC 3115 and M87. 

In many of our simulations, we have found that a typical value for the efficiency of wind production $\eta$ (eq. \ref{eta-efficiency}) is $10^{-3}$. Interestingly enough, this is in good agreement with the mechanical feedback efficiency of $10^{-4}-10^{-3}$ required in cosmological simulations of AGN feedback in the so-called radio mode, in order to offset cooling in galaxy clusters and individual galaxies \citep{Ciotti2010,Sijacki2007,Sijacki2015} and reproduce observations. Therefore, RIAFs could in principle provide efficient feedback to quench star formation in galaxies. 
Given the typical values of $\eta$ found in our simulations, we can use eq. \ref{eta-efficiency} to write
\begin{equation}    \label{power}
\dot{E}_{\rm wind} = 10^{41} \left( \frac{M}{10^8M_\odot} \right) \left( \frac{\dot{M}}{10^{-3} \dot{M}_{\rm Edd}} \right) \ {\rm erg \ s}^{-1}
\end{equation}
where $\dot{M}$ is taken as the accretion rate fed at the outer radius of the accretion flow, as defined previously (cf. section \ref{subsec:eff}). 

We now turn to the comparison of the energetics of our modeled winds with observations of LLAGNs. The "Akira" galaxy hosts a $10^8 M_\odot$ SMBH accreting at $\dot{M} \sim 10^{-4} \dot{M}_{Edd}$ \citep{Cheung2016}. Appying eq. \ref{power} to Akira, we get $\dot{E}_{\rm wind} \sim 10^{40} \ {\rm erg \ s}^{-1}$ which is consistent with the wind kinetic power derived from integral field unit observations of the ionized gas ($\approx 10^{39} \ {\rm erg \ s}^{-1}$; \citealt{Cheung2016}). This wind can inject sufficient energy to offset the cooling rate in both the ionized and cool gas phases in Akira. Moreover, the simple wind model of Cheung et al. gives a constant radially-outward velocity of $310 \ {\rm km \ s}^{-1}$ in a wide-angle cone in Akira. From our simulations, the average velocity of the outflowing particles was $\sim 10^{-3}c \approx 300 \ {\rm km \ s}^{-1}$, which is in excellent agreement with the observations  reported by \cite{Cheung2016}. In conclusion, the properties of the wind observed in the Akira galaxy are well explained as winds from a RIAF as modelled in this work.

The SMBH at the center of Our Galaxy--Sgr A*--is accreting with a Bondi rate of $\dot{M}_{\rm Bondi} \approx 10^{-5} M_{\odot}/yr \approx 10^{-4} \dot{M}_{Edd}$ \citep{Baganoff2003} which taking into account the RIAF solution gives $\dot{M} \sim 0.1 \dot{M}_{\rm Bondi} \approx 10^{-5} \dot{M}_{Edd}$. Using eq. \ref{power} this results in a wind power of $\dot{E}_{wind} = 10^{38} \ {\rm erg \ s}^{-1}$. This estimate is similar to the power previously estimated by different authors \citep{Falcke2000, Merloni2007}. Such winds could be important in explaining the Pevatron observations by the High Energy Stereoscopic System collaboration \citep{HESSCollaboration2016} and the \textit{Fermi} bubbles \citep{Su2010}.

We should note that our winds could be agents of AGN feedback in galaxies hosting SMBHs accreting in the sub-Eddington, RIAF mode. Such feedback would be neither in the radio mode--since it is not through a relativistic jet--nor in the quasar mode--since we are modeling SMBHs accreting at low rates. One class of galaxies which could be subject to this type of feedback--in fact, it seems to be required to explain them--are LLAGNs in the proposed ``red geyser'' mode \citep{Cheung2016,Roy2018}. In red geysers, periodic low-power outflows from the central LLAGN would be able to heat the surrounding gas, prevent any substantial star formation and thereby maintain the quiescence in typical galaxies. The outflows self-consistently modeled in this work can explain the origin of the red geyser mode of AGN feedback.

\subsection{Comparison with previous numerical simulations}\label{subsec:compare-sims}

Our simulations with the ST viscosity, except PL4ST.01, presented the value of $p \sim 0.5-1$, which agrees with the simulations performed by \cite{Stone1999, Yuan2012, Yuan2012b} that had used the same viscosity. The simulations with SS viscosity achieved more efficient winds but with $p \sim 1.1-1.4$, which is slightly below the self-similar, no-wind ADAF solution \citep{Narayan1994}. The resulting power-law dependence of $\rho(r)$ and the fact that $p < 1.5$ are in general agreement with expectations of the ADIOS model \citep{Blandford1999}. It is also in agreement with previous hydrodynamical simulations \citep{Stone1999, Yuan2012b}. However we found very high values of $s \gtrsim 2$ for some simulations, revealing a strong correlation between the radial profile of mass inflow rates and the adopted viscosity parameterization. Considering values of $p$ and $s$, our results for PNST.01 and PNST.1 are the most similar to previous simulations.

On average the efficiency of the winds in our models is in the range $\eta \sim 10^{-3}-10^{-2}$, which is a bit lower than the typical values of $\eta = 0.03$ found by \cite{Sadowski2016} in their GRMHD simulations of RIAFs around nonspinning BHs. We think that the difference is due to the fact that we have not considered magnetic fields in our simulations, which can increase the intensity of outflows due to MHD processes. We intend to investigate the impact of magnetic fields on the outflows in a forthcoming work.

\subsection{Pathologies}

These simulations are purely hydrodynamical, with the angular momentum transport role of the MRI incorporated via an effective viscous stress tensor. MHD effects such as e.g. magnetocentrifugal processes could enhance the production of outflows beyond our estimates in this work. In our simulation the material was ejected via forces created by pressure gradients in the disc --thermally-drive winds. Magnetic fields add into the material a new force component, the Lorentz force, that can enhance the production of outflows and the average energy of the ejected particles. We plan to carry out (GR)MHD simulations to investigate these effects in the future. 

We did not consider the effects of radiation pressure in our simulations, since RIAFs are low-density, optically thin systems, with the radiation field only interacting very weakly with the gas.

Our gravity is represented by the simple pseudo-Newtonian gravitational potential of \cite{Paczynsky1980}. This is clearly not the most accurate description of gravity near the event horizon. Nevertheless, it is a reasonable approximation at larger radii ($r \gtrsim 10 R_S$) and is very useful to keep the calculations conceptually simple (Newtonian) and to save computer time since it avoids the extra computational costs of dealing with metric factors, with the advantage of incorporating the physics of innermost stable circular orbit. For very small radius $r \approx R_S$ our simulation is not very accurate, so we need to restrict our analysis to a slightly larger radius. 

All the simulations were two-dimensional--we assumed complete axisymetry. Three-dimensional simulations could reveal more turbulence in the disc and possible stronger anisotropies in the wind production (e.g. \citealt{Narayan2012}). They are much more computationally expensive, but the upgrade from 2D to 3D can improve the accuracy of the results.

\section{Summary}\label{sec:end}

In this work, we performed two-dimensional numerical, hydrodynamical simulations of radiatively inefficient accretion flows onto nonspinning black holes. Our models were evolved for very long duration of up to $8 \times 10^5 \grt$--comparable to the viscous time of the system. Our initial conditions involved large tori extending up to 500 Schwarzschild radii. Given that the initial conditions of accretion flows are poorly constrained in nature, we explored a diversity of rotation curve profiles and viscosity prescriptions, potentially spanning the diversity of RIAFs that can be found in the centers of galaxies. Our main goal was to investigate the properties of the outflows emanating from these large, hot accretion flows, and compare the properties of these winds with those of low-luminosity AGNs--clarifying along the way their potential for AGN feedback. Here we present a brief summary of our main results: 
\begin{itemize}
\item Our accretion flows produced powerful subrelativistic, thermally-driven winds reaching velocities of up to $0.01 c$.
\item The wind powers correspond to $0.1-1\%$ of the rest-mass energy associated with inflowing gas at large distances, $\dot{E}_{\rm wind} = (0.001-0.01) \dot{M} c^2$, in good agreement with models of the ``radio mode'' of AGN feedback.
\item The properties of our simulated winds are largely in agreement with constraints for the prototypical example of LLAGN wind--the Akira galaxy--and can explain how red geysers are able to heat ambient cooler gas and thereby suppress star formation.
\item Our thermal winds are originated in the corona of the accretion flow ($30\degree \lesssim \theta \lesssim 60\degree$), being produced at distances $\approx 10-100 R_S$ from the SMBH and they can be considered analogous to Parker winds.
\item The equatorial density profile of the accretion flow $\rho(r, z=0)$ displayed a complex behavior which follows the general expectations from the ADIOS models. However, we were unable to make strong statements about the presence of winds based on the indirect information given by $\rho(r)$. 
\item Our models generally displayed a $\dot{M}_{\rm in} \propto r^s$ behavior. However, in some cases the value of $s$ was anomalously high ($s>1$) to be consistent with the expectations of the ADIOS model.
\item Variations in the specific angular momentum profile and the viscosity parameterization caused drastic changes in the accretion flow properties: Even long-run simulations retained some  memory of the initial condition.
\item Most of the winds generated were intermittent with an ``on-off'' behavior. Just a few models displayed continuous winds over the whole simulation time. Sometimes winds were produced in  powerful bursts with $\eta$ reaching close to 100\%. 
\item Thermal winds can remove the excess of angular momentum removal from the accretion flow. Therefore, discs which begin with larger reservoirs of angular momentum will tend to incur more vigorous mass-loss in the winds.
\end{itemize}

We adopted two approaches in analyzing our simulations: (i) looking at the energy and mass fluxes between spherical shells and (ii) using Lagrangian tracer particles to track the wind. The results given by both techniques were consistent with each other, with both approaches supporting the scenario of winds as a generic feature of hot accretion flows. These thermal winds can be a mechanism of feedback in LLAGNs.

We propose two improvements to our simulations: the addition of magnetic fields and improving the dynamical range. Magnetic fields are natural component for accretion flows, since we believe that the mechanism behind the viscosity is the MRI \citep{Balbus1991, Balbus1998}, furthermore mass ejection can be affected by Lorentz force, eventually increasing (or suppressing) the wind strength. 
We need to increase the dynamical range in order to evolve the winds as they flow out of the SMBH's sphere of gravitational influence and into the galactic environment, thereby affecting the host galaxy. These two improvements are the natural next step to the work presented here.

\section*{Acknowledgements}

We acknowledge the help of Francisco Ribacionka who helped us to run our models in the AGUIA cluster. We acknowledge useful discussions with Lu\'is H.S. Kadowaki, Maria Luisa Gubolin, Bhargav Vaidya, Defu Bu, Roderik Overzier, Diego Falceta Gon\c{c}alves and Thaisa Storchi Bergmann. This work was supported by S\~ao Paulo Research Foundation (FAPESP) under grants 2016/24857-6, 2017/01461-2 and 2019/10054-7. This work has made use of the computing facilities of the Laboratory of Astroinformatics (IAG/USP, NAT/Unicsul), whose purchase was made possible by the FAPESP grant 2009/54006-4 and the INCT-A. Research developed with the help of HPC resources provided by the Information Technology Superintendence of the University of S\~ao Paulo. We gratefully acknowledge the support of NVIDIA Corporation with the donation of the Quadro P6000 GPU used for this research.

\bibliographystyle{mnras}
\bibliography{refs}

\appendix

\section{Convergence} \label{app:convergence}

We evaluated the extent to which our models attained inflow equilibrium---i.e. the extent to which they converged---using a similar criterium as the one proposed by \cite{Narayan2012, Sadowski2013}, as follows. We estimated the convergence radius $r_{\rm conv}$ as the radius at which the simulation time $t_{\rm sim}$ equals the viscous time $t_{\rm visc} = r/|v_r|$ $t_{\rm visc}$: 
\begin{equation}
t_{\rm sim} = t_{\rm visc} = \frac{r_{\rm conv}}{|v_r(r_{\rm conv}; t_{\rm sim})|} 
\label{convergence-criteria}
\end{equation}
We restricted our analysis to the equatorial region of the accretion flow and only considered the last $10000 GM/c^3$ of duration of each model. The values of $t_{\rm visc}$ fluctuate as a function of radius making. We took $r_{\rm conv}$ to be the largest value that satisfied equation \eqref{convergence-criteria}. The resulting values of $r_{\rm conv}$ are listed in Table \ref{tab:convergence}.

\begin{table}
\centering
\begin{tabular}{llc}
 \textbf{\#ID} &\textbf{Name}& \textbf{$r_{\rm conv}$} [$R_S$] \\ \hline
00 & PNST.01 & 1024 \\ 
01 & PNST.1  & 280 \\  
02 & PNSS.1  & 5 \\ 
03 & PNSS.3  & 5 \\ 
04 & PNSSV   & 5 \\ 
05 & PL0ST.1 & 201 \\ 
06 & PL0SS.3 & 25 \\ 
07 & PL2SS.1 & 464 \\ 
08 & PL2SS.3 & 1002 \\ 
09 & PL2ST.1 & 444 \\ 
10 & PL4ST.01& 340 \\ \hline 
\end{tabular}
\label{tab:convergence}
\caption{Radius of convergence for all models considered in this work.}
\end{table}

From Table \ref{tab:convergence}, we can see that some of the models  achieved inflow equilibrium out to very large radii, such as models PNST.01 and PL2SS.3 which have $r_{\rm conv} \approx 1000 R_S$. Among our simulations, model PNST.01 is the one with the longest duration whereas model PL2SS.3 displayed the most prominent winds. Models PNSS.1, PNSS.3 and PNSSV  are very similar to each other and displayed poor convergence, as well as PL0SS.3.

Models with higher inflow velocities resulted generally in larger convergence radii. By the same token, the larger velocities are a product of the more efficient angular momentum removal of the ST viscosity prescription, which also led to higher mass accretion rates.

\section{Other simulations} \label{subsec:other}

Besides the two simulation discussed in the main paper, we also performed other simulations (Table \ref{tab:simulations}) which we briefly describe below.

\subsection{PNST.1}

The difference between models PNST.01 and PNST.1 is the adopted value of $\alpha$. In this simulation we use $\alpha = 0.1$ which makes the effects of viscosity more pronounced. In this configuration, the disc loses its original form around $t=60000 \left[ \frac{GM}{c^3} \right]$ and there is no clear outflows in the wind region, following our previous definitions. Simulation PNST.1 becomes very similar to the shape of PNST.01 showed in the left panel of figure \ref{fig:dens-profiles} but in a shorter time and with a 10 times higher accretion rate--as expected for accretion with a higher viscosity.

The values for the equatorial density profile power-law index is shown in Table \ref{tab:allmodels}. The profile indicates the second lowest value of $p$, which indicates mass loss via outflows following the model, but the outflows were not saw and presented a $\eta$ function similar to PNST.01 and the fraction of ejected energy calculated via particles in the system was $\sim 2\%$, a bit below that the value obtained for PNST.01.

\subsection{PNSSV}

This simulation is one of the three simulations with SS viscosity and l(r) following PN profile. This simulation presented extremely low accretion rate and convergence radius. This was reflected with minimal visual changes in torus shape over the simulation evolution. The wind production was intermittent and the material ejection in average was close of the results for PNST.01. However the poor convergence made these results unreliable. The $s$ values for the trio PNSS are unusually high when compared with the literature traditional values, $s \gtrsim 2$, this can be related to lack of convergence in these three simulations.

\subsection{PNSS.1}

This simulation is very close to PNSSV. They share the same specific angular momentum profile and viscosity prescription (SS) with the value of $\alpha$ of both being very close. Therefore, it is not a surprise that the results of models PNSS.1 and PNSSV are very similar. For instance, these two simulations share the same value of $p$ for the equatorial density profile (table \ref{tab:allmodels}), similar $\eta$ variability, ejection rates, launching region and extremely low convergence radius. The effects of the variation of $\alpha(r)$ in the innermost part of the simulation does not change the dynamics of ejection in the wind region and in the outer parts of the accretion disc.

\subsection{PNSS.3}

PNSS.3 is very similar to both PNSS.1 and PNSSV, with the only difference in the choice of $\alpha$. There was not much difference between this simulation and the other two. All members of this trio of models had essentially the same durations ($\sim 400000GM/c^3$), and overall presented very similar results. Including the extremely small convergence radius.

\subsection{PL0ST.1}

This simulation was performed with a constant specific angular moment, $l(R) = {\rm const}$ and a ST-type viscosity profile. It presented a evolution marked by a very strong inflow since the beginning with the material essentially free-falling onto the BH. During the infall, the material piled-up in the inner parts of the disc and formed a spherical accretion flow. We found a jet-like structure arisen in the simulation which has an hydrodynamic origin for the following reason. Material was accreted quite fast due to the strong $\alpha$-viscosity. The disc overfeeds the BH, giving it more than it can take and the accretion becomes spherical. The material were piled up along the polar axis, and the ensuing overpressure creates a vertical structure that looks like a jet. All this process occurred considerably fast, within $15 000 \grt$ after the beginning of the simulation.

Curiously this simulation presented an equatorial density profile $\rho \propto r^{0.97}$, which could indicate existence of outflows. Since we observed only inflows in the model, this confirms that density profiles--taken by themselves--are not a good indicator of the presence of outflows.

The net mass accretion rate in this simulation is essentially the same as PNST.1, even though accretion happened much more rapidly given the larger $\alpha$. Not a single particle escaped to the wind region. The $\eta$ had two bursts along the simulation time with peak of $\sim 0.05$, but most of the time $\dot{E}_{wind} = 0$.

\subsection{PL0SS.3}\label{app:PL0SS3}

PL0SS.3 shares the same $l(R)$ with PL0ST.1, but with other viscosity profile. This one are not similar with the discussed main simulations. The disc shape did not present great changes along the simulation, it maintained its original form during all the $2 \times 10^5 GM/c^3$. The net mass accretion rate here is a bit higher than the rate observed for PNSS.1, PNSS.3, PNSSV and PL2ST.1, and the density profile is similar to the $\rho(r)$ for PL0ST.1, as it had shown in table \ref{tab:allmodels}.

The particles for this simulation presented a behavior slightly similar to the ones from PL0ST.01, the particles have been launched, some were accreted and other followed the external contour of the disc and get ejected near to $\theta = 90 \degree$, hence we do not consider this ejection as a wind. The number fraction of ejected particles in wind region was null. But differently from the other simulations with low value of fraction of ejected energy, the $\eta$ here indicates presence of winds higher than PL2SS.3, which is not consistent with the particle analysis. This come from a diffusion of the huge disc (see panel (a) of figure \ref{fig:initial-conditions}), the disc has diffused and make the calculation of $\dot{E}_{wind}$ unreliable at $r=500R_S$. If we calculate the same integral in a little big radius we can note that $\eta = 0$, different of PL2SS.3. The high value of $\eta$ here is not due to wind production.

\subsection{PL2SS.1}

PL2SS.1 was the simulation with more intense outflows, the fraction of ejected energy is $\sim 20\%$, which is twice the value found for PL2SS.3 in the previous detailed analysis, with a bit smaller time of execution than PL2SS.3. The general aspects of PL2SS.1 were very similar to PL2SS.3, they both shared the same specific angular momentum profile and viscosity prescription, the only difference is the $\alpha$ value, $\alpha^{07} = 0.1$ and $\alpha^{08} = 0.3$. There is minor differences in the density maps between the two simulations, PL2SS.1 showed less ejection in the equatorial plane than PL2SS.3, which was observable in the difference in the slope of density profile from table \ref{tab:allmodels}. The accretion rate and the $\eta$ of these two simulations are very alike.

The main differences between PL2SS.1 and PL2SS.3 are: (i) the net mass accretion rate plot, for PL2SS.3, bottom panel in figure \ref{fig:acc-time}, the net mass accretion rate increased with larger radius, the same is observed for PL2SS.1, with close values, but for PL2SS.1 the net mass accretion rate was outflow dominated, while in PL2SS.3 was inflow dominated. PL2SS.1 is the only simulation in that the mass outflow rate is more intense than mass inflow rate for $30 \lesssim r \lesssim 300R_S$. And (ii) the velocity distribution of the particles, PL2SS.3 velocity histogram, which is showed in the third panel of figure \ref{fig:LP-velocities}, is dominated by particles with $v_r > 0$, for PL2SS.1 there are much more particles with $v_r < 0$, near to the half of the total number. The average velocity of the particles in PL2SS.1 are smaller than PL2SS.3, but is still the second highest average velocity of particles from our sample.

The simulation ejection map was very close to the third panel in figure \ref{fig:LP-map}, both simulations ejected particles from all parts of the disc. PL2SS.1 and PL2SS.3 are similar with each other and very different from the rest of the sample, with some similarities with PL4ST.01.

\subsection{PL2ST.1}

This simulation had the same specific angular momentum profile as PL2SS.1 and PL2SS.3, but with a different viscosity prescription, which led to a complete different result, there was no outflows. The particles had been mostly accreted at high accretion rate, the ejected ones were ejected in the jet region. Like PL0ST.1 in this simulation we had a spherical accretion and the emergence of a jet-like structure formed due to the intense loss of angular momentum of the disc, even with the small running time of $\sim 3.8 \times 10^4 GM/c^3$. There was no winds here. 

The density profile slope was very close to the one found in PNSSV (see table \ref{tab:allmodels}) but they had completely different accretion modes, the torus format evolution have no similarities between these simulations. The ejection fraction and wind efficiency were both null.

\subsection{PL4ST.01}
PL4ST.01 was the only simulation with the original condition $l(R) \propto R^{0.4}$ that did not presented numerical errors in the very first steps of evolution, the implementation of the SS viscosity prescription unfortunately was not possible for this specific angular momentum profile. The results of this simulation were different from all previous setups.

The accretion disc was utterly destroyed in $\sim 1.2 \times 10^5\left[ \frac{GM}{c^3} \right]$ and left some filaments, that looked like a gaseous wig that keep being accreted. The accretion rate decreased after the destruction of the disc, but even with lowered rate it is still orders of magnitude bigger than the accretion rate of the simulations with SS-viscosity (in units of $M_0c^3/GM$). PL4ST.01 had the highest net mass accretion rate, $\dot{m} = \sim 10^{-4} M_0\frac{c^3}{GM}$, of all simulations. 

The fraction of ejected energy from PL4ST.01 was really close to the value of PNSS.3, $\sim 1\%$, but its wind efficiency $\eta$ in the second half of the simulation time is comparable to the value found in PL2SS.3. probably after the torus destruction outflows were produced in PL4ST.01. This scenario is not very physical, because we expect a well-behaved accretion disc that could survive for a long time and not a destroyed disc reduced to some gas filaments in order to explain AGN physics. Another remarkable feature of this simulation is the value of $p=1.53$, which is barely consistent with the assumption of $p<1.5$, considering that we had uncertainties in the calculus.

\section{Winds production}

\begin{figure}
\noindent
\includegraphics[width=\linewidth]{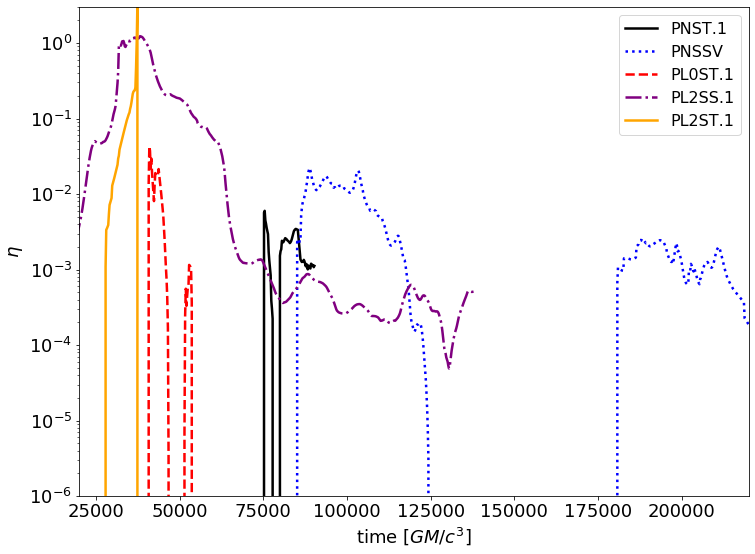}
\caption{Temporal evolution of the wind efficiency $\eta$ as defined in equation \ref{eta-efficiency} for the other simulations. This figure is analogous to \ref{fig:eta-winds}.
}
\label{fig:eta-winds-app}
\end{figure}

The wind production for some simulations can be seen in figure \ref{fig:eta-winds-app}. As we said in section \ref{sec:results}, for various simulations we did not find continuous wind production. Figure \ref{fig:eta-winds} showed non-continuous wind generation for PNST.01. PNST.01 has a wind "flare", the same appeared in figure \ref{fig:eta-winds-app} for PNST.1, PL0ST.1 and PL2ST.1. All these simulations share the same viscosity prescription. 

$7^{\rm th}$ column in table \ref{tab:allmodels} presented the wind activity time for our simulation sample. In average the wind production operates $\sim 50\%$ of the time. As we said in \ref{sec:disc}, outflows similar to ``Parker winds'' can be intermittent \citep{Parker1960}. This is a consistent scenario with our outflows.

\section*{Supporting information}

All simulation data will be made publicly available on figshare upon acceptance of the manuscript for publication.

\bsp	
\label{lastpage}
\end{document}